\documentclass[12pt,preprint]{aastex}

\def \sun{\scriptscriptstyle \odot}
\def \etal{et al.}
\def \kms{km~s$^{-1}$}
\def \numspectra{1139}
\def \numstars{146}
\def \numms{108}
\def \numev{36}
\def \numind{2}
\def \numstarcomp{120}
\def \numstarncomp{26}
\def \totalstars{183}
\def \nonOB{37}
\def \totOB{146}
\def \prOBstars{73}
\def \newOBstars{73}

\def \newOBms{56}

\def \totalmemb{143}
\def \tbleMT{110}

\def \varsystems{36}
\def \pvarsystems{9}

\def \varpercent{30\%}
\def \pvarpercent{42\%}
\def \sbtwos{8}
\def \sb2andsb1{3}
\def \dsbtwos{5}
\def \mt{MT91}
\def \Kndl{Kn00}

\author{Daniel C. Kiminki\altaffilmark{1}, Henry A. Kobulnicky\altaffilmark{1}, 
K. Kinemuchi\altaffilmark{1}, Jennifer S. Irwin\altaffilmark{2}, 
Christopher L. Fryer\altaffilmark{3},\altaffilmark{4}, R. C. Berrington\altaffilmark{1},
B. Uzpen\altaffilmark{1}, Andy J. Monson\altaffilmark{1},
Michael A. Pierce\altaffilmark{1}, S. E. Woosley\altaffilmark{5}}

\altaffiltext{1}{Dept. of Physics \& Astronomy, Univesity of Wyoming, 
Laramie, WY 82070}
\altaffiltext{2}{Department of Astronomy and McDonald Observatory, 
University of Texas, Austin, TX 78712}
\altaffiltext{3}{Theoretical Astrophysics, Los Alamos National Laboratories, 
Los Alamos, NM, 87545}
\altaffiltext{4}{Department of Physics, The University of Arizona,
Tucson, AZ 85721}
\altaffiltext{5}{Department of Astronomy \& Astrophysics, University of California
Santa Cruz, Santa Cruz, CA, 95064}

\begin{document}

\title{A Radial Velocity Survey of the Cygnus OB2 Association}

\begin{abstract} 
  
We conducted a radial velocity survey of the Cygnus OB2 Association
over a 6-year (1999~--~2005) time interval to search for massive close
binaries.  During this time we obtained \numspectra\ spectra on
\numstars\ OB stars to measure mean systemic radial velocities and
radial velocity variations.  We spectroscopically identify
\newOBstars\ new OB stars for the first time, the majority of which
are likely to be Association members.  Spectroscopic evidence is also
presented for a B3Iae classification and temperature class variation
(B3~--~B8) on the order of 1 year for Cygnus OB2 No.~12.  Calculations
of the initial mass function with the current spectroscopic sample
yield $\Gamma = -2.2 \pm 0.1$.  Of the \numstarcomp\ stars with the
most reliable data, \varsystems\ are probable and \pvarsystems\ are
possible single-lined spectroscopic binaries. We also identify 3 new
and 8 candidate double-lined spectroscopic binaries.  These
data imply a lower limit on the massive binary fraction of
\varpercent~--~\pvarpercent.  The calculated velocity dispersion for
Cygnus OB2 is $2.44 \pm 0.07$ \kms, which is typical of open clusters.
No runaway OB stars were found.
\end{abstract}


\section{Introduction}

Cygnus OB2 may be one of the most massive and richest associations in
the Galaxy with $2600 \pm 400$ OB cluster members \citep[hereafter
\Kndl]{Knodl2000} and 90 to 100 O stars \citep[\Kndl]{Come2002}. It
has been studied numerous times due to its richness, proximity (1.7
kpc --- \citet{Han2003,MT91,Shu58,John54}), and high extinction
components with $A_{V} > 5$. The total mass of the cluster is
estimated to be $(4-10) \times 10^{4} \; M_{\sun}$ with a central
stellar density of $\rho_{0} = 40-150 \; M_{\sun}~pc^{-3}$
(\Kndl). The radius of Cyg OB2 has been estimated as large as 30 pc
(\Kndl). For the purposes of this study, we adopt a cluster core
radius of $\sim$15 pc (30\arcmin) inferred from the photometric survey
of \citet[hereafter MT91]{MT91}.

A handful of massive binaries have been discovered in the Association.
Among the OB type binaries are MT421, MT429, MT506, MT554, MT696, Cyg
OB2 No.~5, and Cyg OB2 No.~8a (MT465).  The first four of these are
eclipsing binaries of the Algol type
\citep{PJ98,Kaz2000}. \citet{Rios2004} discovered that MT696 is an
early contact binary (W UMa type) consisting of a late O and an early
B star with a period of 1.46 days. Cyg OB2 No.~5 is possibly a triple
system consisting of an O7Ianfp + Ofpe/WN9 contact binary with period
of 6.6 days
\citep{Leung78} and a B0V star \citep{Ben2001,Wal73,Boh76}. Cygnus OB2
No.~8a is a non-thermal radio emitter \citep{bieg89} and massive
binary system consisting of O6 and O5.5 stars with a period of
$\sim$22 days \citep{Debeck2004}.

Massive close binaries (MCB) are the progenitors of some classes of
energetic phenomena such as supernovae, $\gamma$-ray bursts, and X-ray
binaries \citep{fryer99,fryer98}.  They are also laboratories in which
to study the formation mechanisms for massive stars \citep{bonnell98}.
Direct and indirect evidence reveals that up to 50~--~80\% of massive
stars reside in binary systems
\citep{vanbev2003,gies87,merm2001}. Studies of the MCB frequency are
numerous and include
\citet{Hill2006},\citet{vanbev2003},\citet{zin2003},\citet{merm2001},
\citet{bonnell98},\citet{heuvel84}, and \citet{garmany80}.  Results
have been all but conclusive.  The binary frequency for Galactic
B0~--~B3 stars is at least 32\%, and it may be anywhere between 14\%
and 80\% for OB stars in general \citep{vanbev2003, vanbev98,
merm2001}.  In addition, \citet{vanbev2003} speculates that there may
not be a standard MCB frequency for open clusters and
associations. The long term goal of this study is to provide a
measurement of the MCB frequency, binary star mass ratios and orbital
separations of Cygnus OB2 stars.

Section 2 presents our observations and data reduction. Section 3
discusses the identification of new early type stars and our
derivations of cluster quantities such as the initial mass function,
visual extinction, and distance.  Section 4 describes the calculation
of radial velocities. Section 5 presents the results and a measurement
of the cluster velocity dispersion. Section 6 summarizes the survey
findings.

\section{Observations}
Between 1999 July and 2005 October we observed \totalstars\ stars in
the direction of the Cygnus OB2 Association with the echelle
spectrographs at the Lick 3 m and Keck\footnote{ The W.M. Keck
Observatory is operated as a scientific partnership among the
California Institute of Technology, the University of California and
the National Aeronautics and Space Administration. The Observatory was
made possible by the generous financial support of the W.M. Keck
Foundation.} 10 m telescopes, the Hydra spectrograph at the
WIYN\footnote{The WIYN Observatory is a joint facility of the
University of Wisconsin-Madison, Indiana University, Yale University,
and the National Optical Astronomy Observatory.} 3.5 m telescope, the
Wyoming Infrared Observatory (WIRO) fiber-bundle spectrograph
(WIRO-spec; \citet{pierce2002}) and the WIRO Longslit spectrograph at
the WIRO 2.3 m telescope.  Table~\ref{obs.tab} lists the observing
runs at each facility and the approximate spectral coverages.  Stars
were selected from the UBV photometric and spectroscopic survey of Cyg
OB2 by \mt.  We included \prOBstars\ stars with previous spectroscopic
 OB classification from Table~5 of \mt.  During the initial 1999
Keck run we included \tbleMT\ additional stars from \mt\ having
reddening-free parameter $Q\equiv (U-B) - 0.8 (B-V)$
corresponding to stars earlier than $\sim$B3.
  Our motivation was to obtain a more complete sample of
reddened early type stars.  However, this more inclusive
sample necessarily contained foreground A~--~G stars and background or
foreground OB stars.

At the Lick 3 m telescope, the Hamilton echelle spectrograph was used
to cover the wavelength range $\lambda3620$~\AA\ to $\lambda7675$~\AA\
in 82 spectral orders with a mean spectral resolution of
$R\sim$30,000. These observations occurred over eight nights, 1999 July
21~--~23, 1999 August 21~--~23, and 2000 July 10~--~11. Exposure times
ranged between 240 and 1200 seconds. Hourly exposures of a
Thorium-Argon served to calibrate the wavelength of each exposure to
an RMS of 0.002~\AA (0.12 \kms\ at 5000~\AA). 
The typical resolution was 0.12~\AA\ FWHM at
$\lambda 5700$~\AA\ and 0.08~\AA\ FWHM at $\lambda 3700$~\AA.  A
1.2\arcsec\ $\times$ 2.5\arcsec\ slit decker was used
throughout. Observations of six radial velocity standard stars were
used to confirm the repeatability of the wavelength calibration from
epoch to epoch.

 We acquired spectra at Keck with the HIRES spectrograph
\citep{vogt94} on 1999 July 4-5, 1999 October 18-19, and 2000
September 14-15.  We used the blue collimator to obtain
$R\sim$30,000 spectra over the wavelength range $\lambda 3600 - 5200$
\AA\ ($\lambda3900$ to $\lambda5800$~\AA\ on the 1999 July 4-5
run). On the 1999 July run, spectral regions from $\lambda 5163-5172$
\AA\ and $\lambda 5238-5246$~\AA\ are unusable due to blemishes on the
detector, while on the other Keck runs, spectral regions $\lambda
4512-4517$~\AA\ are unusable. The slit decker C5 measuring
1.15\arcsec\ in the spectral direction and 7.0\arcsec\ in the spatial
direction was used throughout. Periodic exposures of a
Thorium-Argon arc lamp were used to calibrate the radial velocity of
the spectra which were then tied to radial velocity standards.  The
wavelength scale of each exposure is good to an RMS of 0.003~\AA
 (0.18 \kms\ at 5000~\AA). The
output pixel scale is $\sim$0.04~\AA/pix. The mean instrumental
resolution was 0.12~\AA\ FWHM at $\lambda 5000$~\AA\ and 0.09~\AA\
FWHM at $\lambda 3700$~\AA.

At WIYN we used the Hydra spectrograph with the Red camera, 2\arcsec\
blue fibers, and the 1200 l/mm grating in second order to obtain three
1200 s exposures in each of three configurations ($\sim 90$ stars
each).  The spectral coverage was $\lambda3800$ to $\lambda4500$~\AA\
at a mean resolution of $R\sim$4500.  Observations at WIYN took place
over six nights, 2001 August 24, 2001 September 8-9, and 2004 November
28-30. Helium-Neon-Argon lamps were used between each exposure to
calibrate the spectra to an RMS of 0.03~\AA\ (2 \kms\ at 4500~\AA), and
the typical resolution was 1.0~\AA\ FWHM at $\lambda 3900$~\AA\ and
0.82~\AA\ FWHM at $\lambda 4400$~\AA.

At WIRO we used the WIRO-Spec fiber bundle spectrograph with
15$\times$20 fiber array and 1\arcsec\ fibers to achieve $R\sim$4500
over the wavelength range $\lambda 4075 - 4910$~\AA\ at a dispersion
of 0.41~\AA/pix.  Observations at WIRO occurred over eight nights,
2005 July 18-21, and September 18-20, \& 22.  Copper-Argon lamp
exposures were used every 30 minutes to wavelength calibrate the
spectra to an RMS of 0.1-0.2~\AA\ (6.7~--~13 \kms\ at 4500~\AA). The
mean spectral resolution was 0.98~\AA\ FWHM at $\lambda 4500$~\AA.
Typically, three to five 600~s exposures were obtained for each
object.  At WIRO, only a handful of the brightest stars with
previously discovered large-amplitude velocity variations were
observed.

We also obtained spectra of the seven brightest stars exhibiting large
velocity variations with the WIRO longslit spectrograph on 2005
October 13. We used a 2.5\arcsec slit width. The spectra covered the
wavelength range $\lambda3950$ to $\lambda6050$~\AA\ at a dispersion
of 1.12~\AA/pix. Argon lamps were used to calibrate the spectra to an
RMS of 0.15~\AA\ (10 \kms\ at 4500~\AA).  The mean spectral resolution
was 3.2~\AA\ FWHM at $\lambda 4500$~\AA.  Two or three 600 s exposures
were acquired for each object.
      
The data were reduced in IRAF using standard data reduction techniques
which included flat fielding with exposures of an internal or dome
quartz continuum lamp. One dimensional spectra were extracted and
wavelength calibrated by interpolating a wavelength solution
determined from periodic exposures of the arc lamps throughout the
night.  At each observatory and instrument, the variation in the
wavelength solution during the night was small ($<0.5$~\AA),
monotonic, and well constrained by the frequent arc calibration.
Because many of the nights were not photometric, no attempt at flux
calibration was made.  The one dimensional spectra were corrected to a
heliocentric velocity scale by computing the appropriate Doppler
correction for each source, date, time of observation, and observatory
using IRAF tasks RVCOR and DOPCOR. Similar spectra from a given night
were then combined, weighted by the signal to noise.  The final
signal-to-noise ratios vary with wavelength, magnitude, telescope,
instrument, and observing run. They range from 150:1 per pixel near 4400
\AA\ for WIYN observations of the brightest stars to 5:1 for some of
the Lick data where intermittent clouds affected the data.

Observations of the radial velocity standard stars HD012929 (SpT:
K2III), HD171391 (G8III), HD182572 (G8IV), HD187691 (F8V), HR1320
(B2IV), HD1174 (B3V) were obtained on the Lick and Keck nights as a
check on the wavelength calibration. Inspection of the corrected
radial velocity standard star spectra shows excellent agreement
between all epochs, giving confidence that the data are free from
systematic velocity offsets between epochs and telescopes. The
epoch-to-epoch velocity dispersion among velocity standards after
correction to the heliocentric reference frame is $0.05$
\AA\ ($\sim3$ \kms), i.e., less than the instrumental resolution.

The current dataset includes \numspectra\ individual spectra obtained
over 29 epochs for \totalstars\ stars, \totOB\ of which are OB stars.
The mean number of observations per star was 7, the median was 6,
minimum was 1, and the maximum was 19.

\section{New Spectral Classifications and Derived Quantities}
We estimated new spectral types for all stars by visual comparison
with the stellar atlas of \citet{WF90}. Table~\ref{bigtable.tab}
provides a list of all OB stars in our survey using the nomenclature
of \mt.  Columns 2 and 3 list spectral types from the literature
for each object (primarily from \mt; see their Table 5) and as
determined by our survey, respectively. Spectral types were determined
independently by two authors, with the exception of Cygnus OB2 No.~12,
and there was generally good agreement to within one spectral class.
Of the \tbleMT\ stars without previous spectral classifications,
\newOBstars\ were OB stars, while \nonOB\ were A or later.  These
\nonOB\ non-OB stars are listed in Table~\ref{nonOB.tab} and range
from A to M.

\subsection{Cygnus OB2 No.~21}
\begin{figure}
\epsscale{1.0}
\plotone{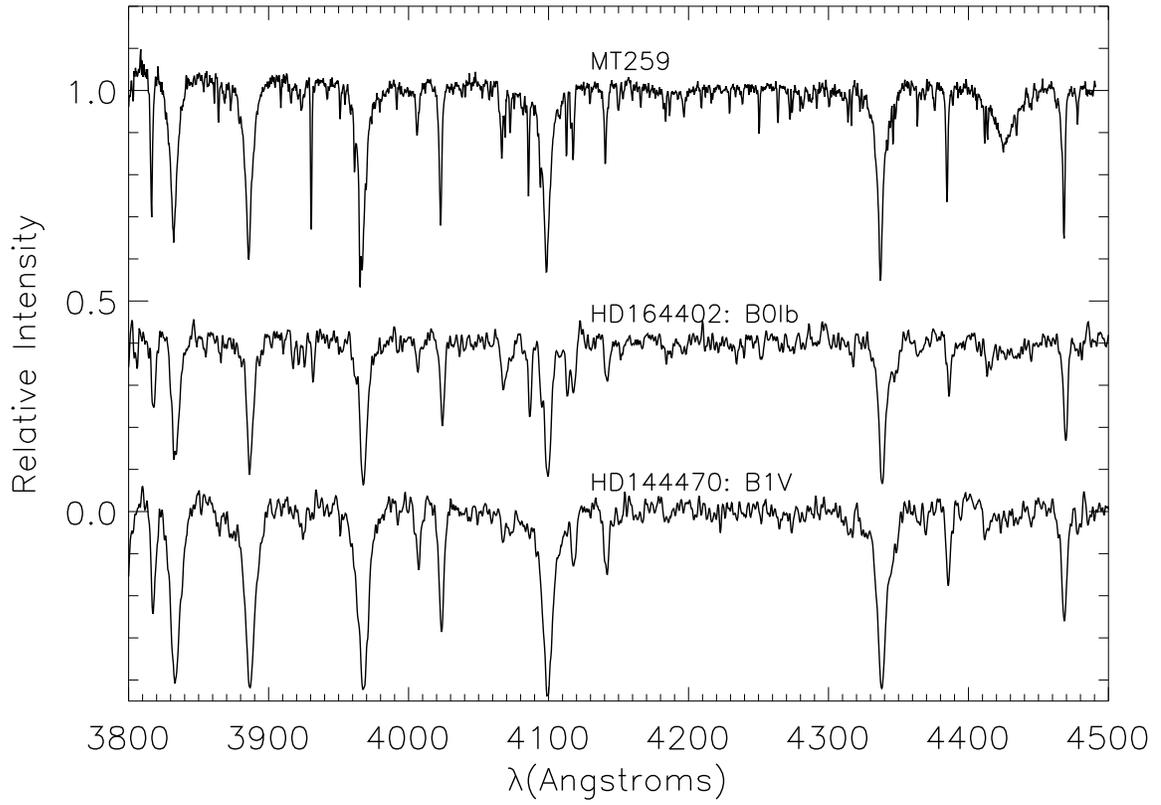}
\caption{2001 September 9 WIYN spectrum of MT259 (Cyg OB2 No.~21) and
comparison spectra from \citet{WF90}. MT259 is more consistent with 
a B0Ib classification. 
\label{MT259} }
\end{figure}

Two thoroughly observed stars, Cyg OB2 No.~12 and No.~21, exhibited 
large discrepancies between our new spectral type and
those listed in the literature. 
\citet{TD91} spectroscopically classify Cygnus OB2 No.~21
(MT259) as a B1 and photometrically as B1V. \mt\ also classified this
star a B1V.  Spectra of all epochs obtained on MT259 were consistent,
and we find that this star is best represented by a B0Ib spectral
classification. In Figure~\ref{MT259} we present a spectrum of MT259
taken on 2001 September 9 at the WIYN observatory. The most
distinguishing feature of this star's spectrum is the large number of
metal lines, especially around H$\delta$ $\lambda$4101~\AA.
Figure~\ref{MT259} also shows the spectra of HD164402 (B0Ib) and
HD144470 (B1V) from the \citet{WF90} spectral atlas for comparison.

\subsection{Cygnus OB2 No.~12}
Cygnus OB No.~12 (MT304) is one of the most notable stars in the
Association. It holds the reputation as one of the most luminous and
reddened objects in the Galaxy \citep{sharp57,john68}. It is also a
bright x-ray source \citep{harn79} and variable radio emitter
\citep{gun2003}.  The X-ray emission is best explained by wind driven
shocks \citep{wald2004}, where No.~12 has a measured high stellar wind
of $V_{WIND} = 1400$ \kms\ \citep{leit82,bieg89}, and an expanding
shell of $V = 3100$ \kms\ \citep{WZ2003}. It has been shown to vary
photometrically \citep{Gott78,Voel75}, and \citet{SL80} submit that a
binary companion may provide an explanation for this. \mt\ suggest the
possibility that No.~12 is a Luminous Blue Variable because it
appears to vary spectroscopically as well as photometrically and has
an extremely high luminosity.

\begin{figure}
\epsscale{1.0}
\plotone{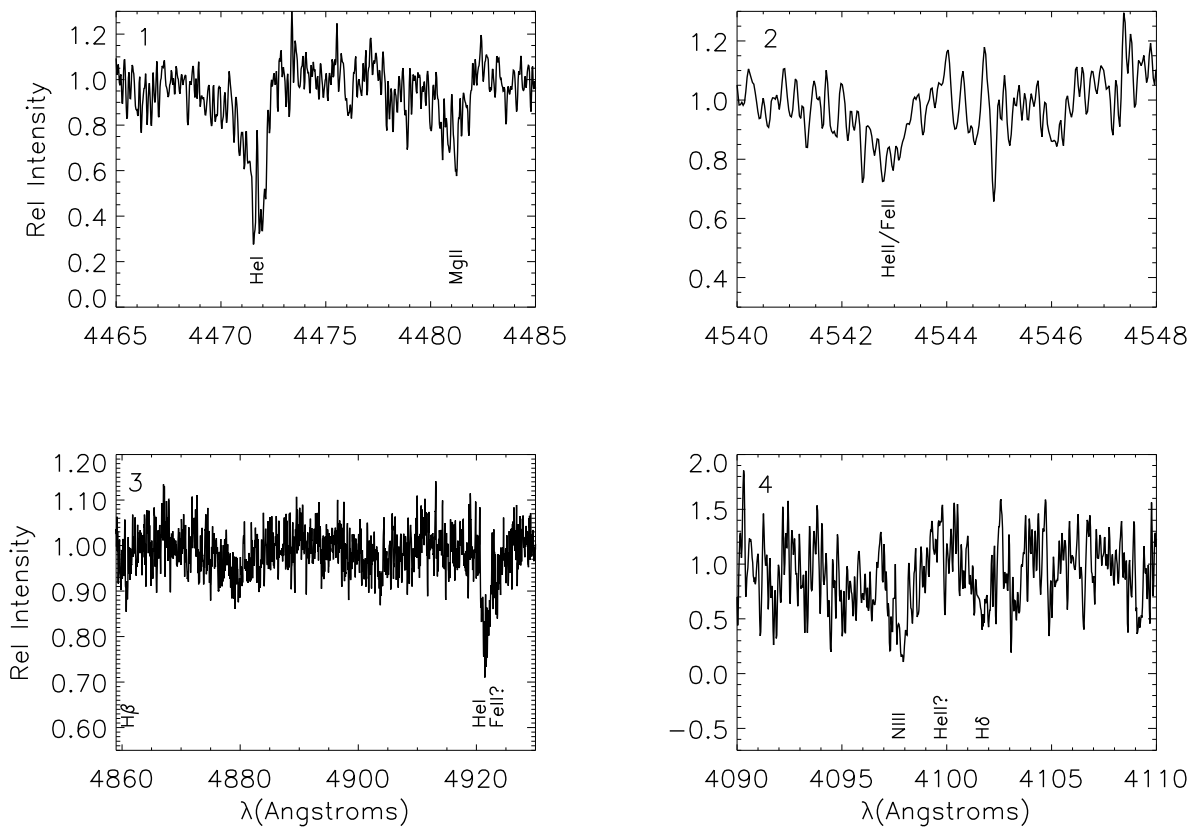}
\caption{A close up of portions of MT304 (Cyg OB2 No.~12) from the 2000 September 18
 spectrum taken at Keck. Panel one shows the important HeI
$\lambda$4471 to \ion{Mg}{2} $\lambda$4481~\AA\ ratio of approximately 2:1. Panel
two shows a \ion{He}{2}/\ion{Fe}{2} blend at $\lambda$4542~\AA, very weak in early B
stars, and absent in later types. Panel three shows \ion{He}{1} $\lambda$4922~\AA\
and an H$\beta$ line filled in by emission. The strength of $\lambda$4922~\AA\
and the weak or absent \ion{Fe}{2} indicate a type of B5 or
earlier. Panel four shows an emission-filled H$\delta$, \ion{N}{3}
$\lambda$4097~\AA\ and possible \ion{He}{2} $\lambda$4100~\AA. \label{no12}}
\end{figure}

Most studies adopt a spectral type for No.~12 of B5Iae (\mt) or B8Iae
\citep{SL80}.  \citet{abbott81} adopt B3Iae but little justification
for such an early classification is found in the literature. We find
evidence for a B3Iae spectral type and a temperature class variation
between B3 and B8 by comparison with the OB star atlas of \cite{WF90},
the supergiant atlas of \citet{Lenn92}, and the spectral atlas of
\citet{Yama77}.  Figure~\ref{no12} presents segments of the spectrum
obtained on 2000 September at Keck. Panel 1 shows the
temperature-sensitive ratio \ion{He}{1} $\lambda$4471~\AA\ to
\ion{Mg}{2} $\lambda$4481~\AA, the strongest evidence for a B3Iae
spectral type. \ion{Mg}{2} is a weaker feature than \ion{He}{1} for
temperature classes earlier than B7 and is stronger than \ion{He}{1}
for a B8 class or later. The observed ratio is approximately 2:1,
characteristic of a B3 temperature class. It should be noted that the
spectrum presented by \citet{SL80} shows a ratio characteristic of a
B8 class \citep[Figure~1]{SL80}, and the spectrum presented by \mt\
shows a ratio characteristic of B3 (although it is reported as B5 due
to additional arguments; see Figure 12 in \mt).  The presence of the
absorption feature $\lambda$4542~\AA\ (Panel 2) provides additional
evidence for a B3 classification.  The relative intensity and
asymmetric profile of the absorption suggests a blend of the two lines
\ion{He}{2} $\lambda$4542~\AA\ and \ion{Fe}{2} $\lambda$4542~\AA. This
is generally only seen in the earliest of B stars. Panel 3 shows
H$\beta$ $\lambda$4861~\AA, \ion{He}{1} $\lambda$4922~\AA, and a weak
\ion{Fe}{2} $\lambda$4923. The strength of the \ion{He}{1} absorption
in relation to the weak \ion{Fe}{2} absorption suggests an earlier
type.  Panel 4 shows an example of deep absorption in one of the
higher ionization lines, \ion{N}{3} $\lambda$4097~\AA. There is also a
possible \ion{He}{2} feature at $\lambda$4100~\AA, next to H$\delta$
$\lambda$4101~\AA.  We see additional evidence for weak \ion{He}{2}
absorption toward the blue end of the spectrum (i.e.,
$\lambda$3710~\AA\ and $\lambda$3720~\AA). The absence of \ion{Si}{4}
$\lambda$4089~\AA\ suggests a classification no earlier than
B3. Panels 1 and 4 also display clear examples of documented
emission-line cores in No.~12 (\ion{He}{1} $\lambda$4471~\AA\ and
\ion{N}{3} $\lambda$4097~\AA).

\begin{figure}
\epsscale{1.0}
\plotone{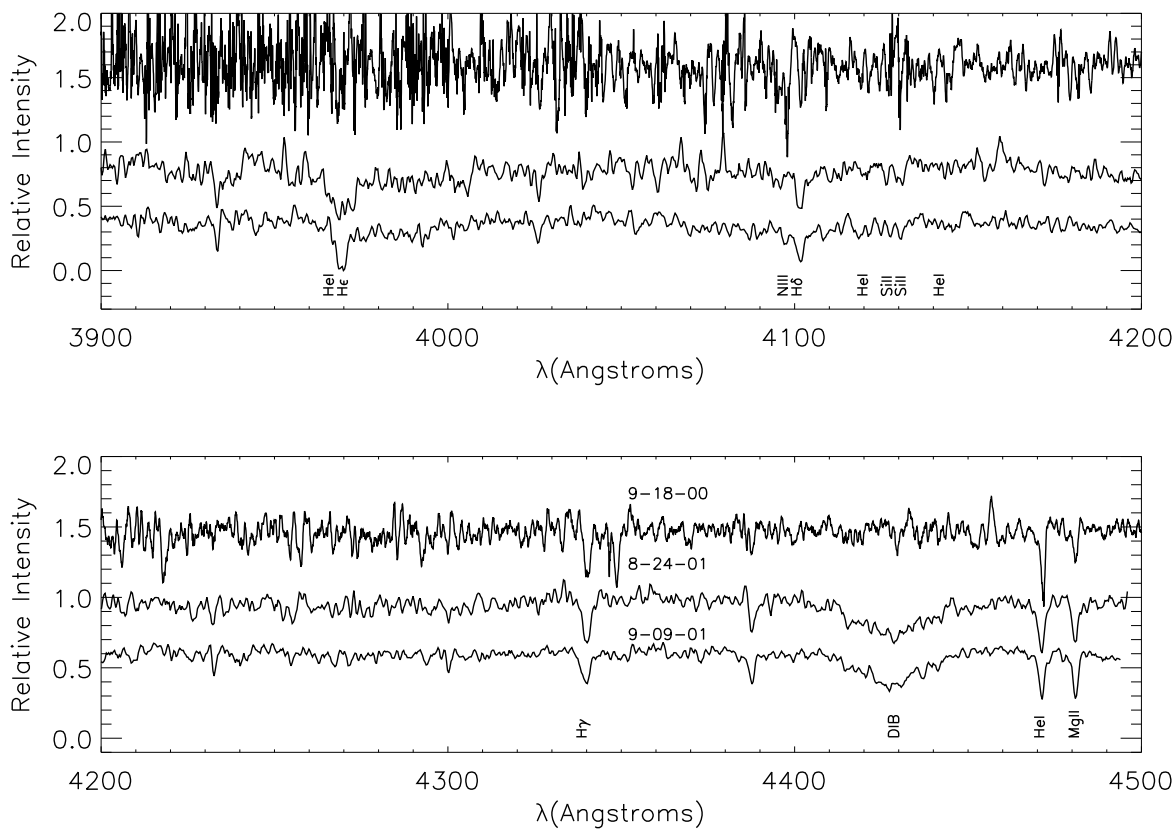}
\caption{A comparison of the spectra for MT304 (Cyg OB2 No.~12)
obtained over one year: Keck 2000 September 18 (smoothed by 3~\AA)
followed by WIYN 2001 August \& September. Changes in Balmer line
equivalent width, the \ion{N}{3} $\lambda$4097 depth, and the \ion{He}{1}
$\lambda4471$~\AA\ to MgI $\lambda4481$~\AA\ ratios are also shown.
The sequence shows an evolution from B3I in 2000 September to at least
B8I in 2001 September.  \label{no12b}}
\end{figure}

Fluctuations in the spectrum of No.~12 are seen in our 2001 August and
September spectra obtained at WIYN and are shown in
Figure~\ref{no12b}. \citet{SL80} noted small spectral fluctuations in
their spectra over a time scale of days.  Figure~\ref{no12b}
shows three of our spectra obtained over the span of one year and
demonstrates a temperature class evolution between B3 and B8.  The
upper spectrum is from Keck HIRES on 2000 September 18, and the lower
two spectra are from WIYN Hydra on 2001 August 24 and September 9.
The broad interstellar band feature near 4428~\AA\ is absent in the
high-resolution Keck spectrum because it has been fitted and removed
during continuum normalization. The ratio of \ion{He}{1} $\lambda4471$
\AA\ to \ion{Mg}{2} $\lambda$4481~\AA\ changes from over 2:1 to near
1:1. The strength of the \ion{Si}{2} $\lambda$4128~\AA\ \&
$\lambda$4130~\AA\ doublet becomes stronger than the adjacent
\ion{He}{1} lines. \ion{N}{3} $\lambda$4097~\AA\ becomes noticeably
weaker than H$\delta$ $\lambda$4101~\AA\ and \ion{Fe}{2} $\lambda$4232
\AA\ becomes much stronger. In addition, the Balmer line strength also
increases. Emission in the Balmer and higher ionization lines is also
less dominant in the 2001 August and September spectra which are best
described by a classification of B6 and B8. It is more dominant in the
2000 September spectrum which is best described by a classification of
B3. A sign of late B classification is the appearance of the
\ion{Mg}{1} doublet at $\lambda$5173~\AA\ and $\lambda$5184~\AA,
however the August and September WIYN spectra did not cover this
section of the spectrum and the Keck spectrum showed no evidence of
them.  The Lick data did cover this spectral range but were not of
sufficient S/N for detailed classification.

\subsection{Visual Extinction and Distance}
Table~\ref{bigtable.tab} provides the calculated visual extinctions
(Column 6) and distance moduli (column 13) for the \numstars\ OB stars
in the direction of Cyg OB2. The values were calculated from apparent
visual magnitudes (Column 7), absolute visual magnitudes (Column 8),
colors, (\bv) (Column 9), and intrinsic colors,
$\mathrm{(\bv)}_{\mathrm{0}}$ (Column 10).  The apparent visual
magnitudes and $(B-V)$ colors were obtained from \mt. The absolute
visual luminosities were adopted from \citet{FM05} for the O stars and
\citet{HM84} for the B stars. The intrinsic
$\mathrm{(\bv)}_{\mathrm{0}}$ colors were based on adopted
spectral types and obtained from \citet{Weg94}.

\begin{figure}
\epsscale{1.0}
\plotone{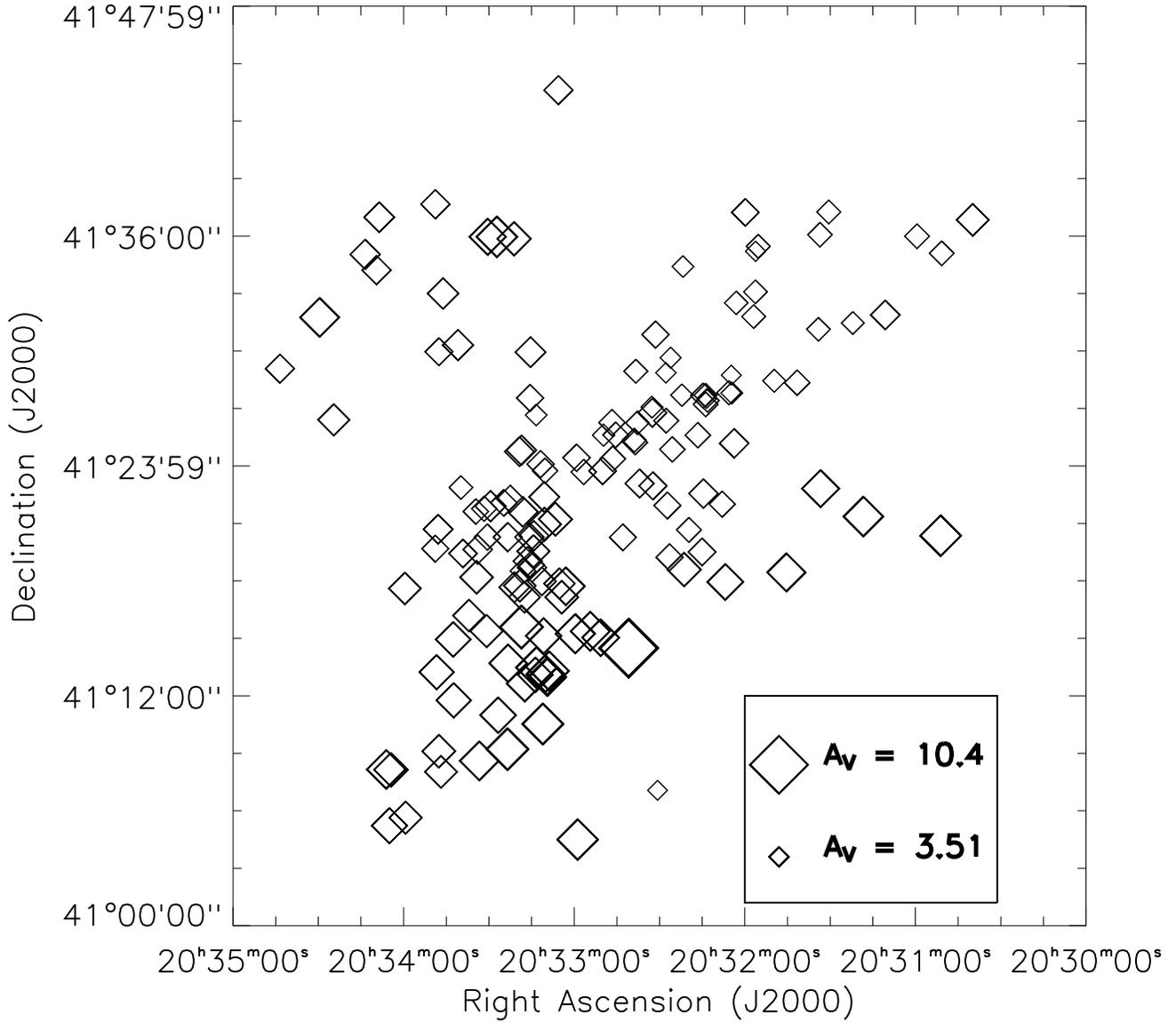}
\caption{A map of calculated extinction, $A_V$, for \numstars\ OB
stars in the direction of Cygnus OB2.  The relative size of the symbol
is proportional to $A_V$. \label{av}}
\end{figure}

\begin{figure}
\epsscale{1.0}
\plotone{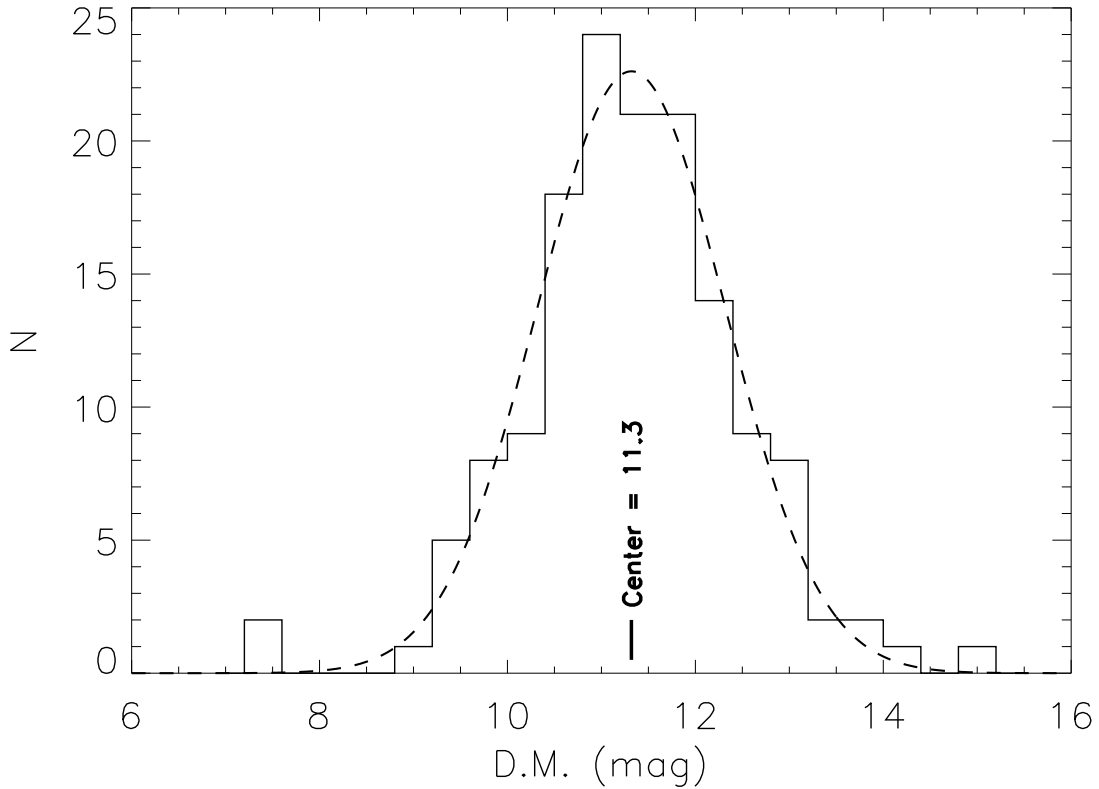}
\caption{Histogram of spectrophotometric distances to the \numstars\
OB stars in the direction of Cygnus OB2.  The mean distance modulus is
11.3 magnitudes or $\sim$1.8 kpc. The approximately Gaussian dispersion
is due to uncertainties in the absolute magnitudes, particularly of
the post-main-sequence stars. We adopt all stars with distance moduli
between 8.5 and 14.5
as provisional members of the association while
acknowledging that this generous
range includes some foreground and background objects.
\label{disthist} }
\end{figure}

We computed the visual extinctions, $A_V$, using

\begin{equation}
A_{V} = R_{V}[(\bv) - (\bv)_0],
\end{equation}

\noindent where $R_{V}$ is the ratio of total to selective
extinction. We adopt $R_{V}=3.0$ based on the studies of
\cite{Han2003} and \mt. Visual extinctions range between $A_V=3.5$ and
$A_V=10.4$ (for No.~12) with a mean near $A_V=5.4$ mag, consistent
with prior results (e.g., \mt).  Figure~\ref{av} shows an extinction map
for all OB stars in the direction of Cygnus OB2 where
the relative symbol size is proportional to $A_V$.
Using these calculated extinctions, we computed photometric distance
moduli to all of the OB stars using

\begin{equation}
DM = V - M_{V} - A_{V},
\end{equation}

\noindent where $V$ is the apparent visual magnitude and $M_{V}$ is
the absolute visual magnitude. Figure~\ref{disthist} shows a histogram
of the computed distance moduli for all \numstars\ OB stars in the
direction of Cyg OB2. This histogram peaks at $D.~M.\simeq11.3$
magnitudes or $\simeq1.8$ kpc, in good agreement with the commonly
adopted distance estimate of 1.7 kpc \citep{Han2003,MT91,Shu58}.  The
distribution is approximately Gaussian with a width of $\sigma$=1.0
mag. The large width reflects the uncertainties on the absolute visual
magnitudes, particularly for evolved stars, and uncertainties in
temperature and luminosity classification.  \citet{Han2003} found a
similar scatter in DM of up to 1.5 mag and mean distance moduli from
10.08~--~11.16 mag, depending on the adopted absolute magnitude
calibrations.  An additional systematic uncertainty arises from the
presence of unresolved binaries which lead to smaller inferred distance
moduli.  Unresolved binaries are at least partially responsible for
the asymmetry in the distribution in Figure~\ref{disthist}. A third
source of uncertainty might also stem from the ratio of total to
selective extinction, $R_{V}$. \citet{Patri2003} found an average
$R_{V}=3.1$ for the Cygnus region, but a spread of $\sim1$. A spread
of 1 in $R_{V}$ translates to a spread in $A_{V}$ of approximately
1~--~2 magnitudes for our sample. The compound extinction components in
combination with the density of documented O stars in Cygnus OB2 could
create conditions under which the dust grain size distribution 
and composition vary and lead to 
a variable $R_{V}$ across the cluster.

\subsection{New Early Types and Cluster Membership}
The true distance moduli shown in Figure~\ref{disthist}
are listed in column 13 of Table~\ref{bigtable.tab}. For the purposes
of this study, we accept as provisional members all stars with
distance moduli between $\sim$8.5~--~14.5 mag which lie within the broad
Gaussian distribution. Three stars fall outside of the distribution
(MT427, MT573, and MT304). It will be shown in the following section
that including all such stars does not significantly affect the
computed slope of the initial mass function. About half of these were
previously identified spectroscopically as probable members by
\mt. The other probable members were determined photometrically as a
group by \mt\ but not named individually. The total sample consists of
\numms\ main sequence stars, \numev\ evolved stars and \numind\ with
indeterminate luminosity class.  We have identified \newOBstars\ early
type stars without previous spectral classification in the literature.
Newly classified early type stars are denoted in column 14 of
Table~\ref{bigtable.tab} with an 'n'.  Of these, \newOBms\ are main
sequence stars, mostly type B0 and later.

As a secondary means of assessing membership for the early type stars,
we examined the equivalent widths of the diffuse interstellar band
(DIB) absorption feature at 4428~\AA.  We found that nearly all of the
newly identified stars had $EW_{DIB}=3\pm1$~\AA, consistent with the
established association members \citep{snow2002}.  Given the large
spread in reddening and DIB equivalent width for Cyg OB2, and the fact
that \citet{snow2002} found a poor correlation between reddening and
DIB strength for this cluster, we conclude that this relation is not a
useful discriminant of members from field stars.

\subsection{Initial Mass Function}

Studies of the slope of the CygOB2 initial mass function have reached
disparate conclusions. \Kndl\ used JHK 2MASS Point Source Catalog
\citep{2MASS} photometry and a K-band mass-luminosity relation to
calculate a mass function slope of $\Gamma=-1.6\pm0.1$. This is
steeper than previous studies of Cyg OB2. \mt\ utilized the
evolutionary models of \citet{Maeder88} with an H-R diagram
constructed from their spectroscopy and ``best'' photometry to obtain
a slope of $\Gamma = -1.0 \pm 0.1$.  This is relatively shallow
compared to the canonical \citet{salpeter} value of $\Gamma=-1.35$.
\citet{Mas95} found a similar slope of $\Gamma=-0.9 \pm 0.2$, using
the same technique over a mass range of $15 \; M_{\sun} - 25 \;
M_{\sun}$. In both calculations, a coeval system was assumed.

\begin{figure}
\epsscale{1.0}
\plotone{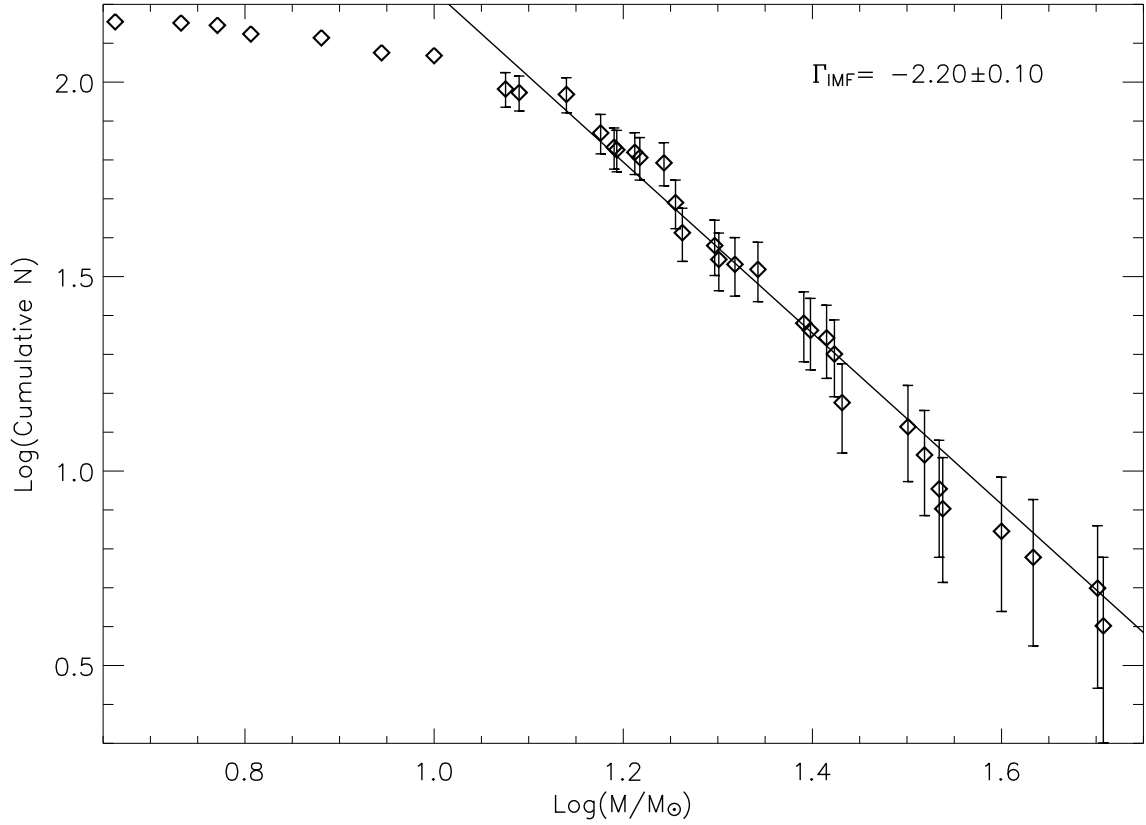}
\caption{A plot of mass distribution for provisional members of Cyg OB2.
The initial mass function is calculated using a cumulative mass count. 
The solid line is a linear fit to masses greater than
$log(M/M{\sun})=1.0$. A slope of $\Gamma = -2.2 \pm 0.1$ has been
calculated. Limiting the analysis to stars
having distance moduli within 1.5 mag of the mean
does not change the slope by more than 0.04. \label{imf}}
\end{figure}

We calculated an IMF slope using the spectroscopic masses for the
previously classified \prOBstars\ OB stars and the \newOBstars\ newly
classified OB stars reported here.  Present-day masses for each star
were taken from \citet{FM05} for the O stars and interpolated from the
tables of \citet{drilling} for the B stars. Initial masses were also
estimated for each of the evolved stars from the \citet{schaerer}
stellar evolutionary models. Table~\ref{bigtable.tab} lists the
present-day and adopted initial masses for each star in Columns 11 and
12.  Figure~\ref{imf} shows the cumulative logarithmic mass
distribution of \totalmemb\ Cyg OB2 stars with
spectroscopic masses (minus Cygnus OB2 No.~12 owing to the uncertainty
in its spectral type, mass loss rate, and current mass).  We use a
cumulative rather than a differential mass distribution to measure the
IMF slope in order to mitigate uncertainties caused by discrete mass
bins and the choice of bin size and placement.  Diamonds indicate the
logarithmic number of stars with a logarithmic mass greater than or
equal to that point. The error bars reflect Poisson statistics. It is
clear from the change in slope at $log(mass)\simeq1.0$ that the
spectroscopic survey becomes incomplete at masses below $\sim15$
$M\sun$ corresponding approximately to a B1V star.  The solid line
represents a linear fit to all points greater than this cutoff.  We
obtain a slope of $\Gamma=-2.2 \pm 0.1$. This is much steeper than
the canonical \citet{salpeter} value of $\Gamma=-1.35$.  This value is
also steeper than the results of \Kndl\ and significantly more so than
\mt. The possible inclusion of foreground or background stars may bias
the slope toward steeper values. However, the IMF slope does not vary
by more than $0.04$ from the nominal value of $\Gamma=-2.20$ when we
remove stars which lie farther than $1.5\sigma$ (24 stars) from the
mean distance modulus of $11.3$.  The dominant explanation for the
difference between our IMF and the previous results is a systematic
difference in the predicted initial stellar masses between the models
of \citet[used by \mt]{Maeder88} and \citet[used in this
study]{schaerer}.  The implied masses of \citet{schaerer} are
systematically lower for a given bolometric luminosity and effective
temperature compared to the models of \citet{Maeder88}.  Effectively
this means that we have fewer high-mass stars than \mt\ which results
in a steeper slope.  We also calculated the slope of the present-day
mass function (PDMF) to be $\Gamma =-2.3 \pm 0.1$, which is more
appropriate for comparison to the photometric PDMF results of
\Kndl. It should also be noted that our sample is composed only of
stars within the core of Cygnus OB2 while the \Kndl\ sample
encompasses a large number of stars outside the core region.

\section{Calculating the Relative Radial Velocities}

We used two methods to calculate the relative and absolute radial
velocities: Gaussian profile fitting and cross-correlation.  For
early type stars which have a small number of broad spectral features,
profile fitting can be a reasonable approach.  This method entailed
fitting Gaussian profiles to the strong H and He absorption lines
($H\beta$, $H\gamma$, $H\delta$, \ion{He}{1} $\lambda$4471,and \ion{He}{1} 
$\lambda$4388).  We used a five-parameter Gaussian\footnote{The IDL
package ``MPFIT'' written by Craig B. Markwardt, NASA/GSFC Code 662, is
substituted for the standard ``GAUSSFIT'' and ``CURVEFIT'' due to its
greater control over the fitting parameters.} of the form

\begin{equation}
f(x) = A_{0}e^{-\frac{1}{2}(\frac{x - A_{1}}{A_{2}})^2}+ A_{4}x+ A_{3}
\end{equation}

\noindent where $A_{0}$ is the \textit{Depth}, $A_{1}$ is the
\textit{Center}, $A_{2}$ is the \textit{FWHM}, $A_{3}$ is the
\textit{Constant Term}, and $A_{4}$ is the \textit{Linear Term}.  The
code performs initial fits to the H$\beta$ and He~I $\lambda 4471$
lines of a template spectrum (usually the highest quality Keck
spectrum) in order to measure the widths of the H and He lines,
respectively.  The width parameter, $A_2$, for the H and He lines is
subsequently held constant at these initial values.  The line centers
are fixed relative to each other using the rest wavelengths from
atomic line lists \citep{cowley}.  The code then fits profiles to all
of the H and He lines with $A_0$, $A_3$, $A_4$ as free parameters for
each line. The best-fit parameters are then saved and used to
construct a ``template'' consisting of 5 Gaussian components.  This
template is then stepped through a range of velocities from -160 \kms\
to 160 \kms\ at a stepsize of 1 \kms\ to minimize the $\chi^2$ between
the template and spectra from each epoch.  At each velocity step,
$A_0$, $A_3$, and $A_4$ of each component are allowed to vary
separately to achieve the best fit.  The velocity of the global
$\chi^2$ minimum is adopted as the most probable velocity for each
epoch. Because the H lines are stronger than than the He lines, they
were given more weight in the calculation of the total $\chi^2$.
Spectra from WIYN have limited wavelength coverage and do not include
$H\beta$, so only 4 Gaussian components are fit. Spectra from WIRO
have limited wavelength coverage and do not include $H\gamma$, hence,
only 4 Gaussian components are fit.  Velocity uncertainties were
calculated using the $\Delta\chi^2$ statistic \citep{numerical}.

Gaussian fitting proved simple to automate for a large dataset.  We
also found it to be more robust for low signal-to-noise spectra and in
regions at the edges of spectral orders in the echelle data.  However,
it often yielded larger uncertainties.  We found that in the case of
poor quality spectra the fitting routine would provide solutions for
local or no minima.

We also measured radial velocities with cross-correlation techniques
using the IRAF/XCSAO task which is part of the RVSAO package
\citep{xcsao}.  This method had the advantage of using information in
all spectral features, although for main sequence stars most of the
cross-correlation power stems from the few strong H and He lines.  As
the template for cross-correlation we tried using both an observed
spectrum (usually a high-quality Keck spectrum) and a model stellar
atmosphere \citep[TLUSTY]{LHub2003} of the appropriate effective
temperature and gravity.  The model atmosphere is not rotationally
broadened to match individual stars in our sample, but in the vast
majority of cases the model line widths appear well-matched to the
spectra.  In most cases, the analysis using atmospheric model
templates produced smaller velocity uncertainties than either the
results from Gaussian fitting or cross-correlation with observed
spectral templates.  The smaller uncertainties result from the higher
signal-to-noise of the model templates and the fact that this method
uses the power of many additional spectral features in the templates
and data to constrain radial velocities.  Varying the gravity and
effective temperature of the template produced little or no change in
the correlation results.  Therefore, we adopt the velocities obtained
by cross-correlation with the model atmospheres.
Table~\ref{electr.tab} contains the star name, heliocentric Julian
date, relative radial velocities, and 1$\sigma$ error estimates for
each epoch of observation in columns 1~--~4 respectively. The
uncertainties are calculated within XCSAO using the equation,

\begin{equation}
\sigma_v = \frac{3w}{8(1+r)},
\end{equation}

\noindent where $w$ is the FWHM of the correlation peak and $r$ is the 
ratio of the correlation peak height to the amplitude of antisymmetric 
noise \citep{xcsao}.

\begin{figure}
\epsscale{1.0}
\plotone{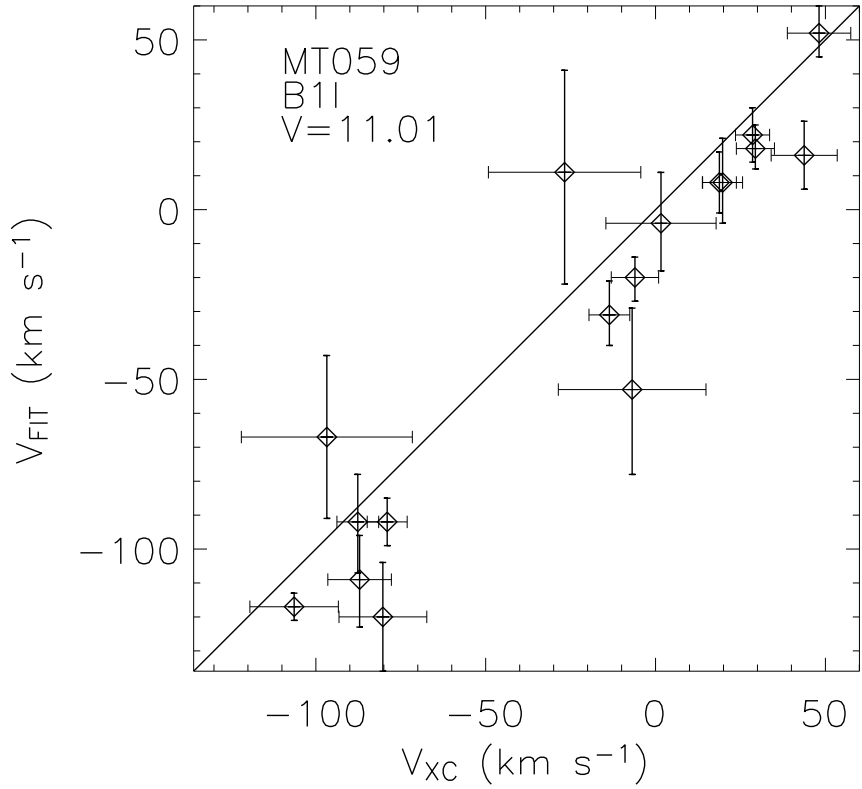}
\caption{A comparison of relative radial velocity results for Gaussian
profile fitting versus cross-correlation techniques for MT059. A minor
offset of $\sim$10~--~15 \kms\ is observed in most comparisons, most
likely attributed to the use of a model atmosphere as a template for
the cross-correlation. On average, the larger errors belong to the
Gaussian profile fitting.
 \label{59comp}}
\end{figure}

\begin{figure}
\epsscale{1.0}
\plotone{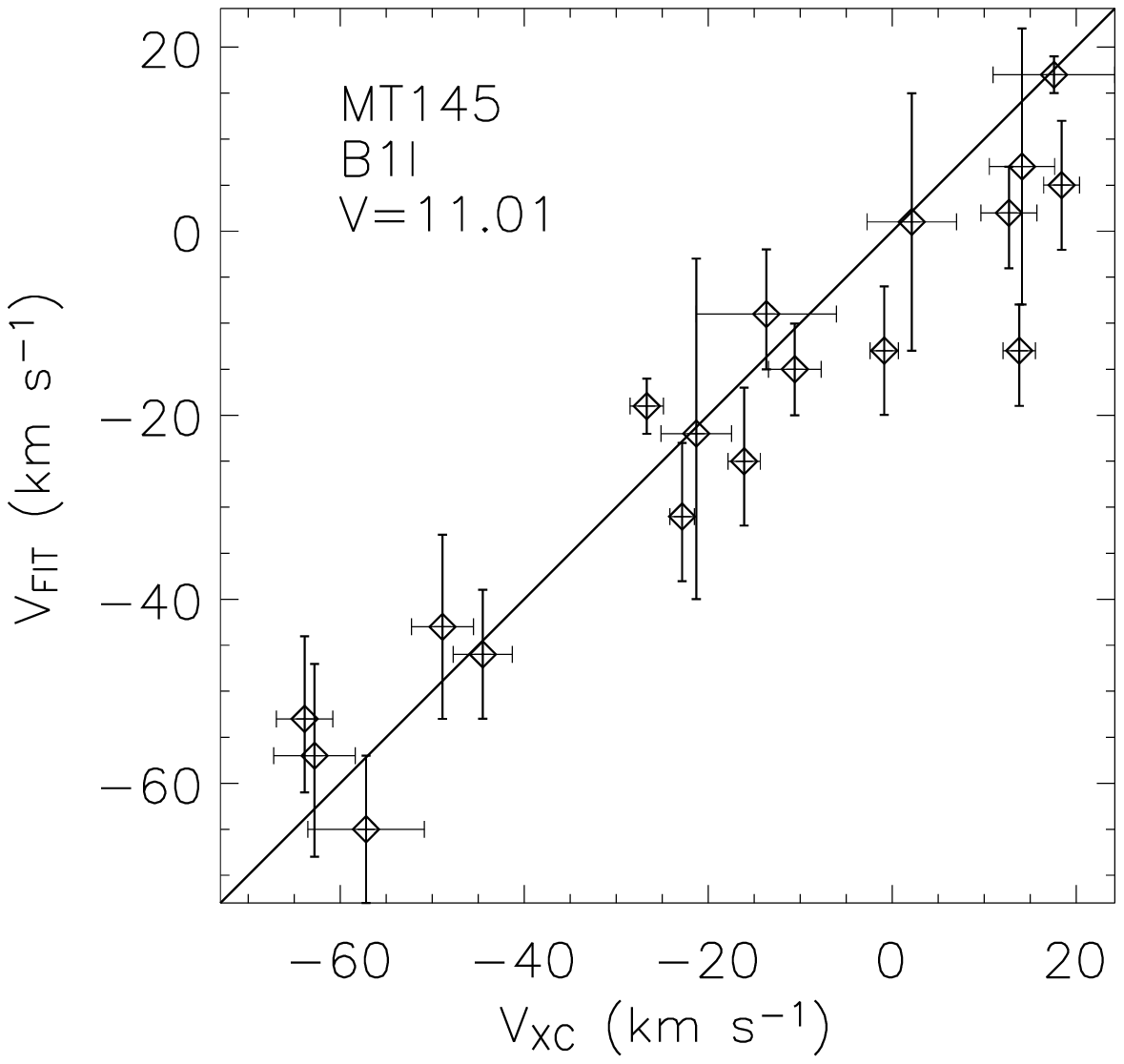}
\caption{A comparison of relative radial velocity results for Gaussian
profile fitting and cross-correlation techniques for
MT145.  \label{145comp}}
\end{figure}

\begin{figure}
\epsscale{1.0}
\plotone{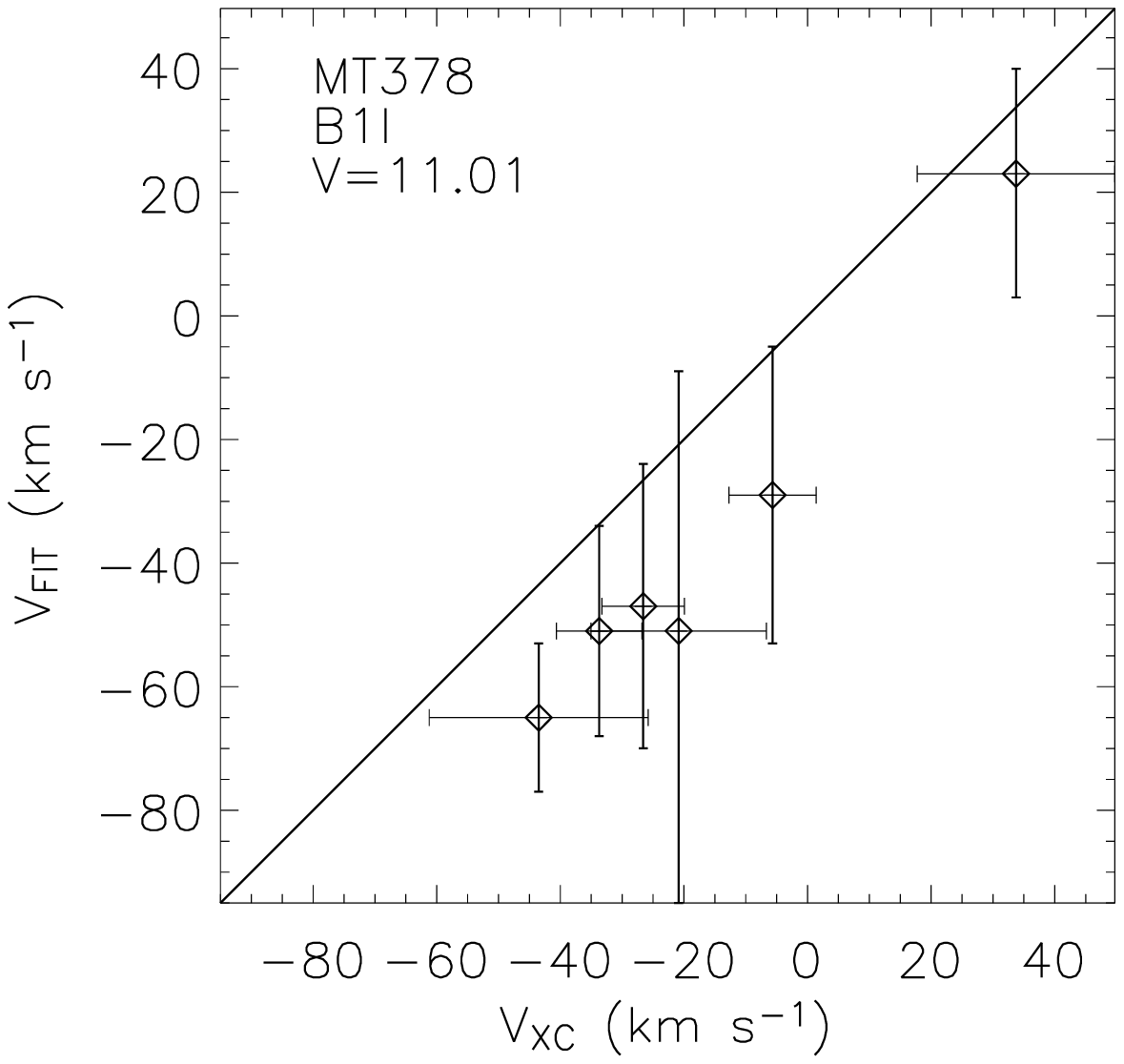}
\caption{A comparison of relative radial velocity results for Gaussian
profile fitting and cross-correlation techniques for
MT378.  \label{378comp}}
\end{figure}

\begin{figure}
\epsscale{1.0}
\plotone{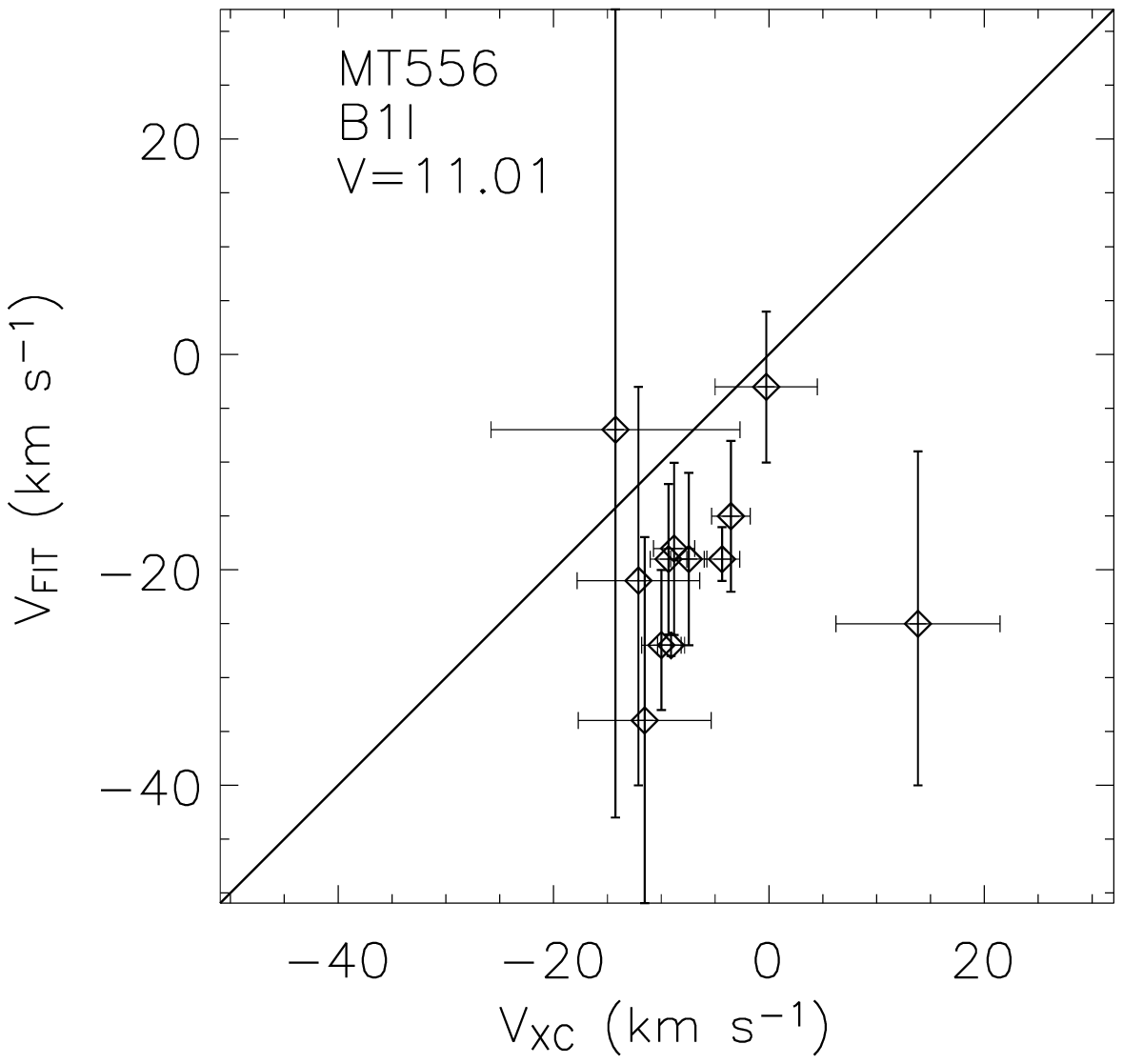}
\caption{A comparison of relative radial velocity results for Gaussian
profile fitting and cross-correlation techniques for
MT556. \label{556comp}}
\end{figure}

We found that the resultant relative radial velocities and their
uncertainties showed good agreement between the two analysis methods.
Figures~\ref{59comp} through \ref{556comp} show comparisons of
velocities and their uncertainties for an O8V, O9III, B0V, and B1I star
respectively.  These figures demonstrate that there is a strong
correlation between velocities obtained with the fitting versus the
cross-correlation techniques. Frequently, the error bars are smaller
for the cross-correlation analysis, especially for evolved stars which
have more spectral features that are utilized by cross-correlation
methods but not by our Gaussian fitting code.  Deviations from the 1:1
correspondence are generally consistent with the shown error,
indicating that the statistical uncertainties are well-estimated.  In most
cases, there is also a zero-point offset of $\sim10-15$ \kms\ in the
sense that the profile-fitted velocities are smaller (i.e., more
negative) than the results from cross-correlation.  The magnitude of
this offset is similar for both main-sequence and evolved stars.  Such
a modest offset is not unexpected given that the line centers of the
model atmospheres should not necessarily agree with the rest
wavelengths of the five H and He lines used by our profile fitting
code.
 
From the initial sample of \numstars\ possible Cyg OB2 stars, we
discarded \numstarncomp\ stars from the sample that had fewer than 3
observations, poor quality spectra or emission lines that interfered
with the cross-correlation analysis.  The stars showing emission were
discarded because the phenomenon might be time-variable and produce
apparent radial velocity variations that mimic the effects of orbital
motion. Furthermore, we eliminated a small number ($\sim20$ out of
1139) individual spectra with low signal-to-noise (mostly Lick data
from a run plagued by clouds).  The remaining sample used for velocity
analysis consists of \numstarcomp\ stars.

For the remaining \numstarcomp\ objects, Table~\ref{veltable.tab} lists
the identification for each star based on \mt\ notation (column 1),
the spectral type (column 2), the number of observations for each star
(column 3), $V_{avg}$, the weighted average heliocentric velocity
(column 4), $V_{mid}\equiv 0.5 (V_{max} + V_{min})$, the average of
the largest and smallest observed heliocentric velocities (column 5),
$V_h=0.5 (V_{max} - V_{min})$, a measure of the velocity
semi-amplitude (column 6), $V_{rms}$, the velocity dispersion (column
7), and $\overline{\sigma}_v$, the mean velocity uncertainty (column
8).  $V_{avg}$ and $V_{mid}$ are both measures of the systemic
velocity for the star, although both are susceptible in different ways
to sampling effects and measurement errors.  $V_{avg}$ would provide a
robust measure of the true systemic velocity if the observations
uniformly sample all orbital phases in a binary system with zero
eccentricity.  However, our sparse sampling coupled with the
likelihood that some orbits are eccentric renders this measure less
than ideal. $V_{mid}$ provides an alternative measure of the systemic
velocity, which is less susceptible to under sampling but may be more
prone to measurement uncertainty.  We use both $V_{avg}$ and $V_{mid}$
as indicators of the systemic velocities, and we find that they yield
comparable results.  $V_h$ and $V_{rms}$ are both measures of the
observed velocity variations for a given stellar system.  $V_h$ is a
measure of the velocity semi-amplitude of the system, albeit an
imperfect one, when the velocity curve is not sampled at all phases.
$V_{rms}$ is another measurement that reflects the level of velocity
variations in a system.  We use $\overline{\sigma}_v$ to describe a
characteristic uncertainty averaged over all observations.

There is an additional source of random and/or systematic uncertainty
which may contribute to the overall error budget of each measurement.
Stellar photospheric line profile variations may be present among some
of the most massive stars, especially the evolved stars in our sample
(31 of \numstarcomp\ stars or about $\sim$26\% are post-main-sequence stars).
Line profile variations attributed to atmospheric pulsations are
observed in $\geq$77\% of evolved O stars and in some Be stars
\citep{penrod,vogt83} but rarely among dwarf stars
\citep{fullerton96}.  These phenomena could mimic the effects of bona
fide orbital velocity variability. Irregular variability due to these
effects might also mask a true low amplitude binary. More frequent
time sampling of candidate variables is the only way to identify bona
fide binaries and reduce mis-identification.

\section{Results}

\subsection{Velocity Variations}

\begin{figure}
\epsscale{1.0}
\plotone{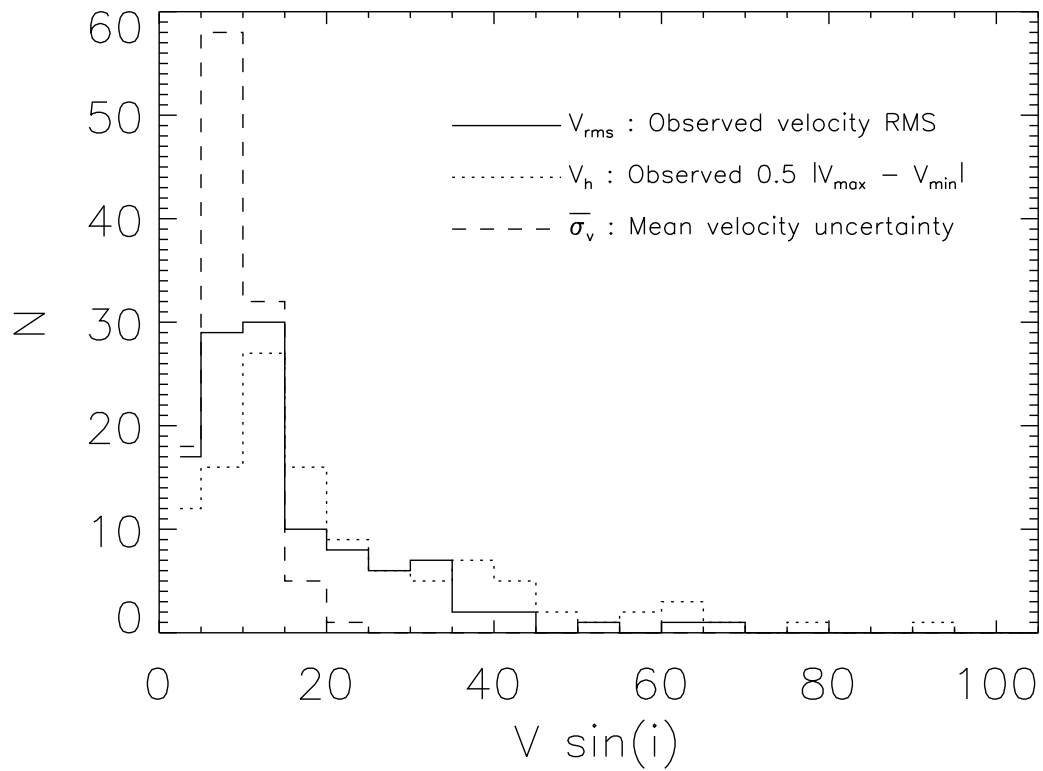}
\caption{The distribution of observed velocity dispersions, $V_{rms}$
(solid line), velocity semi-amplitudes, $V_h\equiv
0.5|V_{max}-V_{min}|$ (dotted line), and the mean velocity
uncertainties (dashed line) for the sample.
    \label{chip}}
\end{figure}

Figure~\ref{chip} shows a histogram of the observed radial velocities
and mean uncertainties.  This Figure illustrates the distribution of
velocity dispersions, $V_{rms}$, calculated from the multiple
measurements of the \numstarcomp\ OB stars (solid line) along with the
distribution of mean velocity uncertainties, $\overline{\sigma}_v$
(dashed line).  The dotted line shows the distribution of $V_h$.  The
lowest velocity bin from 0 to 5 \kms\ is sparsely populated because
observational errors scatter the data into higher velocity bins.  The
maximum observed semi-amplitudes fall mostly between 10 and 40 \kms,
with a significant tail toward higher velocities out to $\sim$90 \kms.
The uncertainties lie in the characteristic range 5~--~15 \kms.

\begin{figure}
\epsscale{1.0}
\plotone{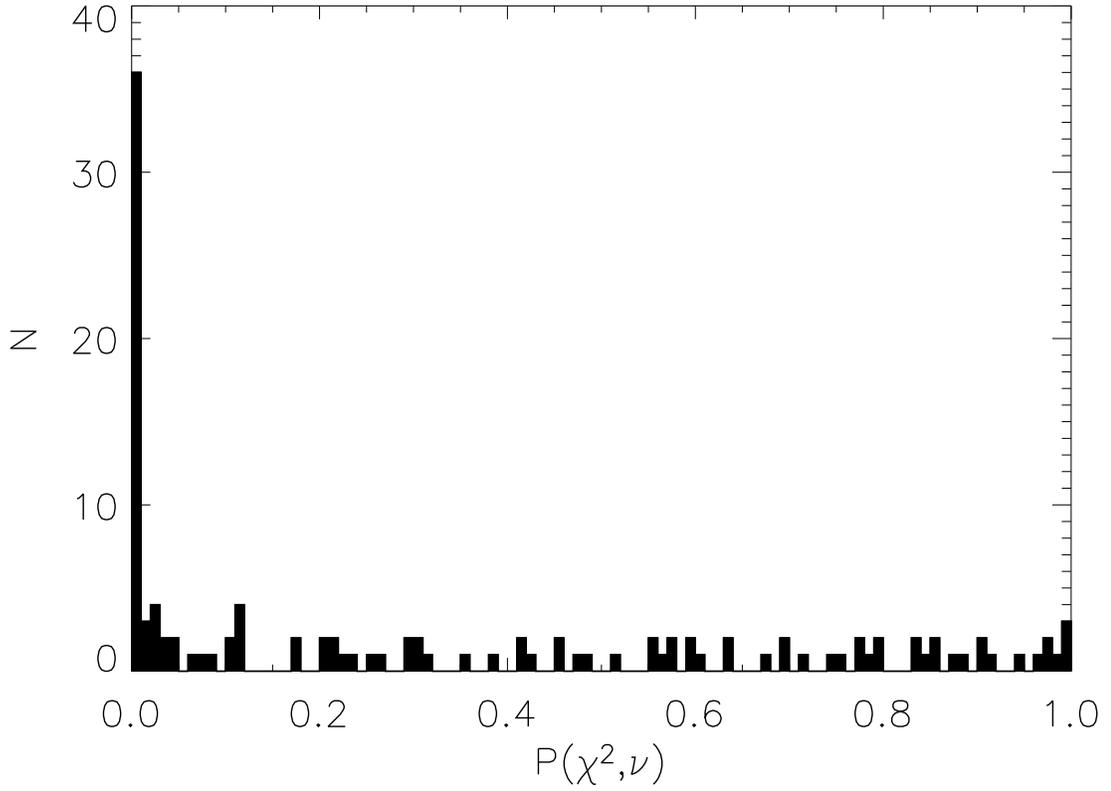}
\caption{The distribution of probabilities that $\chi^2$ (as determined
about the weighted mean) would be exceeded given $\nu=N_{obs} - 1$ degrees
of freedom. The discontinuity at $P(\chi^2,\nu) = 0.01$ shows the 
change in the distribution from probable non-variables to probable 
variables within the sample. \label{Pchi}}
\end{figure}

The velocity results in Table~\ref{veltable.tab} show a wide range of
characteristics, including stars with large variations and stars with
no significant variations. We used the method presented in
\citet{DM91} to identify probable spectroscopic variables in our
survey. In Figure~\ref{Pchi} we show the distribution of probabilities
that $\chi^2$ (as considered about the weighted mean of the
measurements) would be exceeded given $\nu=N_{obs} - 1$ degrees of
freedom. As in \citet{DM91}, the distribution is relatively flat
except for values $P(\chi^2,\nu)\leq 0.01$.  Column~9 of
Table~\ref{veltable.tab} lists the computed probabilities for the
\numstarcomp stars examined. We identify in column~10
all stars with $P(\chi^2,\nu)\leq 0.01$ as probable SB1s (\varsystems\
stars). In addition, we also identify the stars with
$0.01<P(\chi^2,\nu)<0.04$ as possible SB1s (\pvarsystems\ stars). The
remaining objects are probable systems comprised of single stars,
systems viewed at low inclinations, systems with very low mass
companions or very long periods.

\begin{figure}
\epsscale{1.0}
\plotone{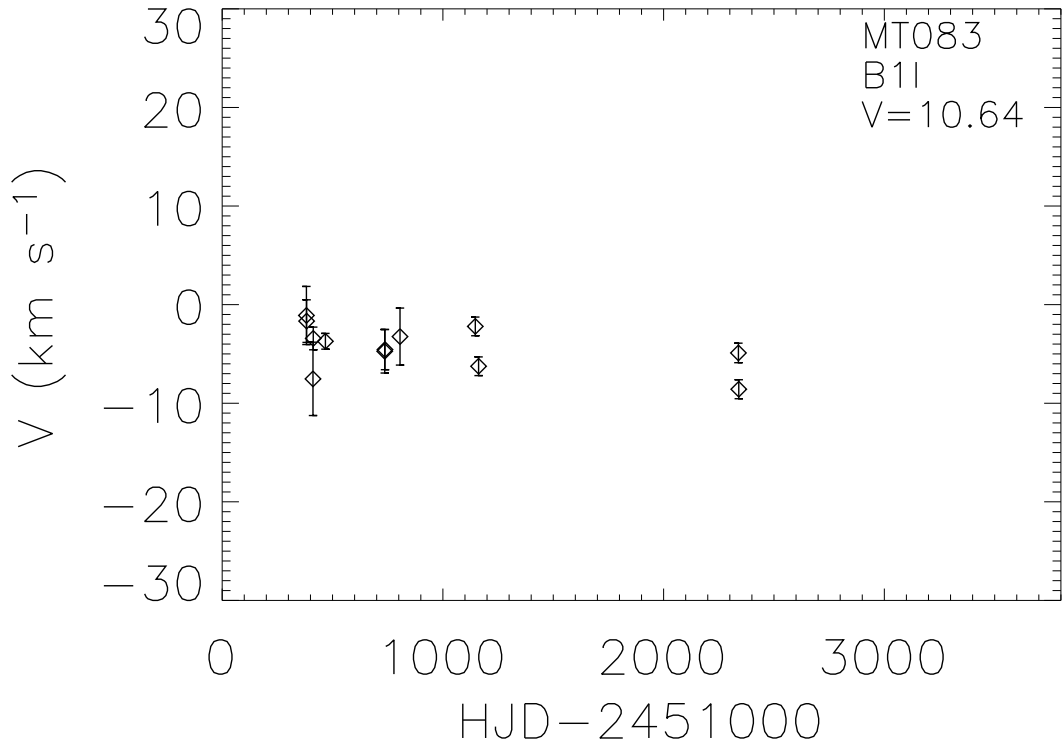}
\caption{Relative radial velocity variation for MT083. MT083 shows variations
$\leq5$ \kms\ but is considered an SB1 since
$P(\chi^2,\nu)<0.0004$
   \label{83novar}}
\end{figure}

\begin{figure}
\epsscale{1.0} 
\plotone{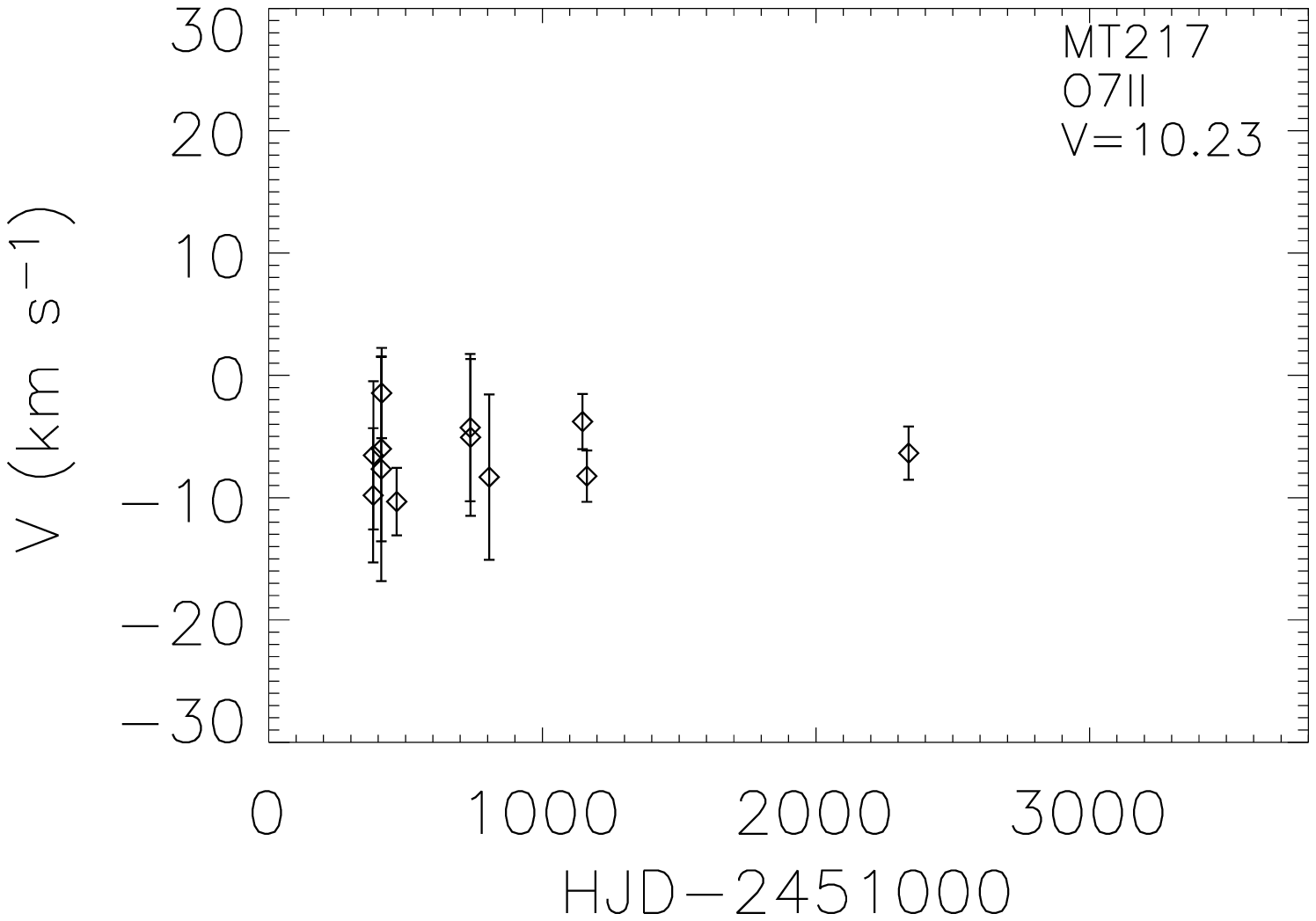}
\caption{Relative radial velocity variation for MT217, a system with
little or no variation.  \label{217novar}}
\end{figure}

\begin{figure}
\epsscale{1.0}
\plotone{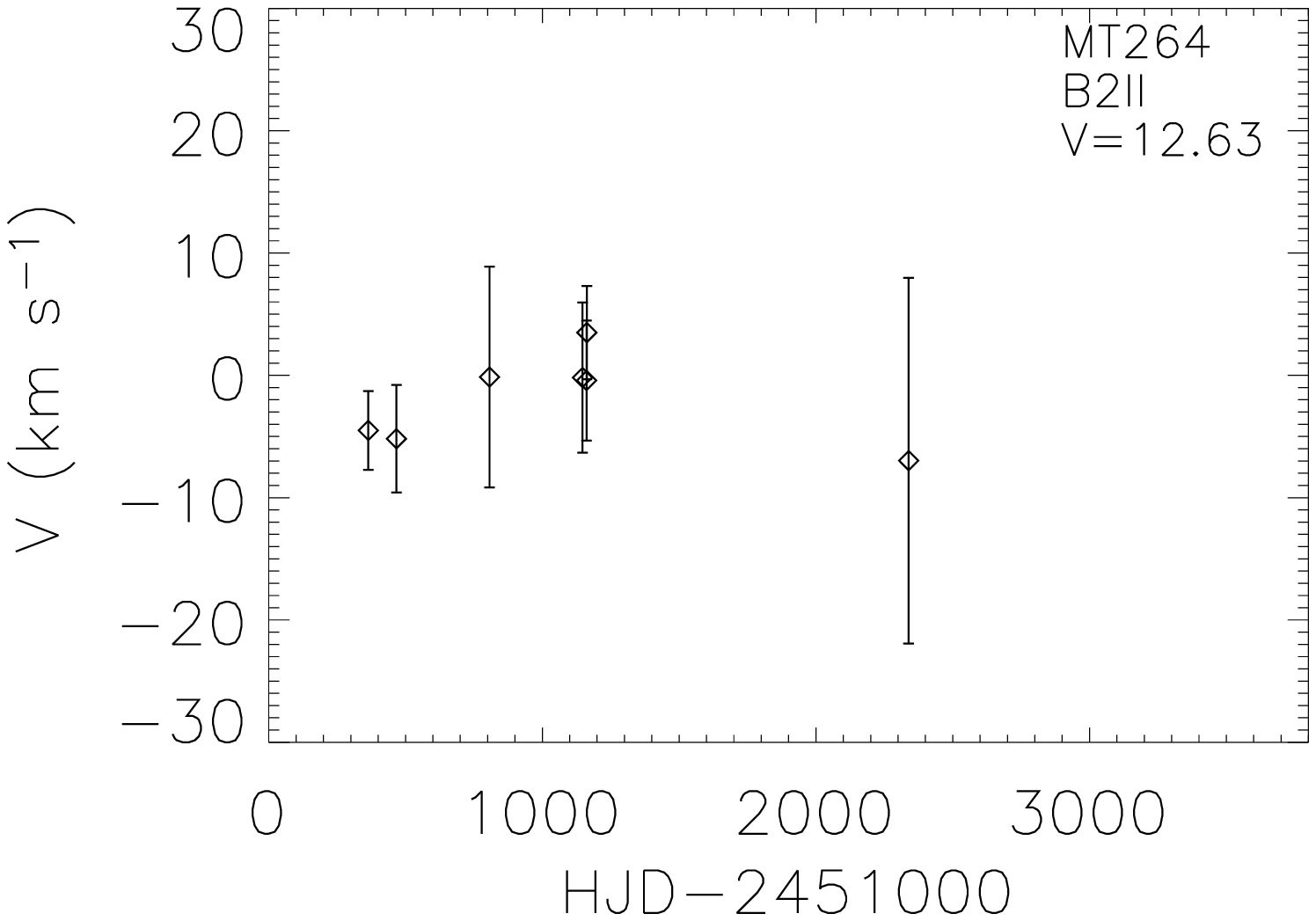}
\caption{Relative radial velocity variation for MT264, a system with little or no
variation.   \label{264novar}}
\end{figure}

\begin{figure}
\epsscale{1.0}
\plotone{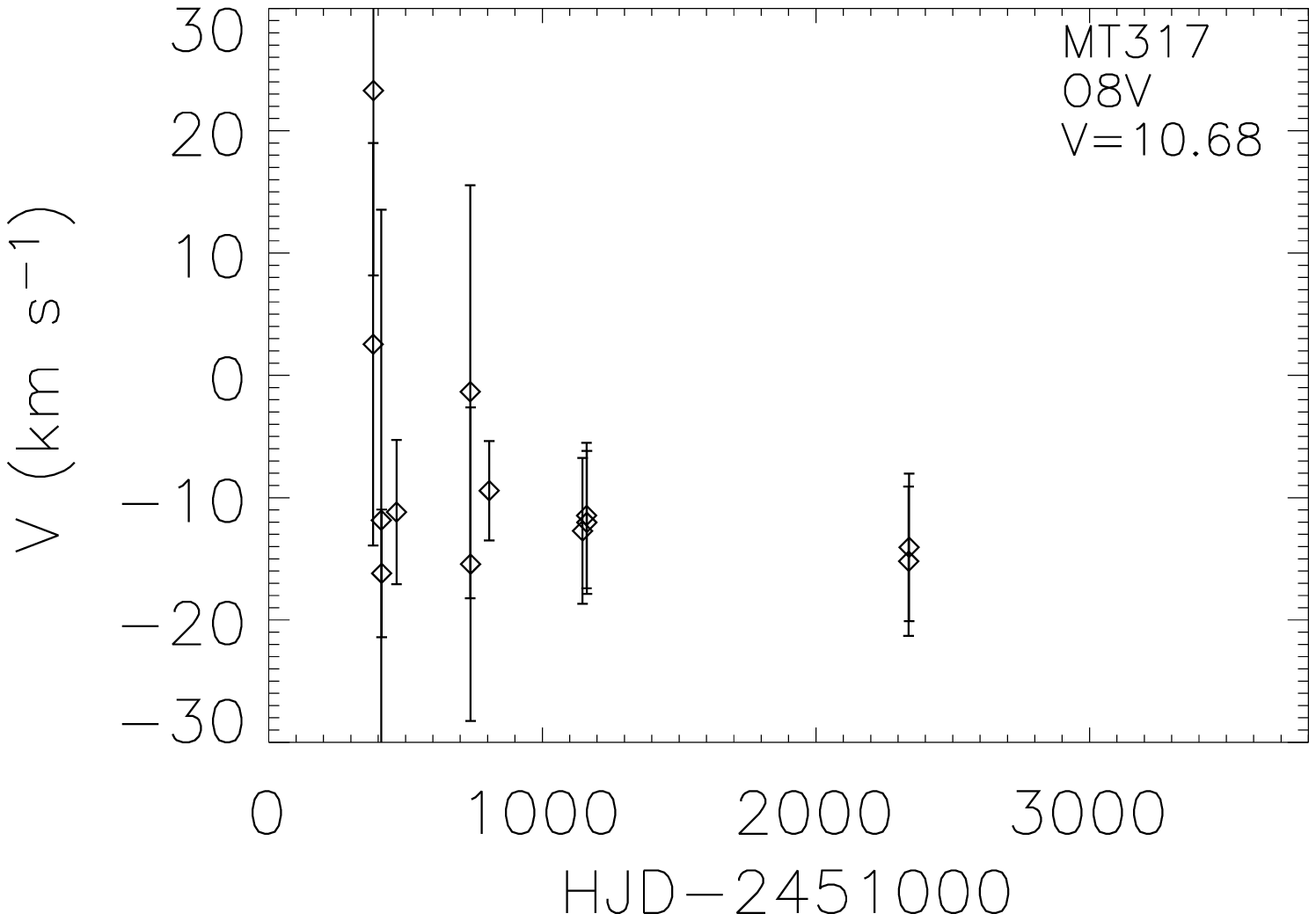}
\caption{Relative radial velocity variation for MT317, a system with little or no
variation.   \label{317novar}}
\end{figure}

In Figures~\ref{83novar} through \ref{317novar} we show heliocentric
velocity versus time plots for four objects with minor or no velocity
variations: MT083 (B1I; SB1), MT217 (O7II), MT264 (B2II), and MT317
(O8V).  The data for most objects cover a time interval of $\sim5.5$
years from 1999 July through 2004 November.  The evolved systems among
this group have some of the smallest velocity uncertainties in the
survey ($<5$ \kms) and are particularly well suited to assess the
level of possible systematic velocity errors from epoch to epoch.
MT083 in particular (Figure~\ref{83novar}) exhibits no epoch-to-epoch
variations greater than $\sim5$ \kms\ but has $P(\chi^2,\nu)<0.0004$,
indicating that it is a low-amplitude spectroscopic variable.  This
system also exhibits some photometric variability, raising the
possibility that the velocity variations may be attributable to
atmospheric activity \citep{kukar81}.  In comparison with the small
velocity variations present among these four objects, the magnitude of
the Doppler correction due to the Earth's orbital motion ranges from
$\sim8$ \kms\ in July to $\sim -16$ \kms\ in November.  The constancy
of the velocities, especially for evolved stars which have small
uncertainties, among this subset of objects provides reassurance that
systematic errors do not dominate the results.

\begin{figure}
\epsscale{1.0}
\plotone{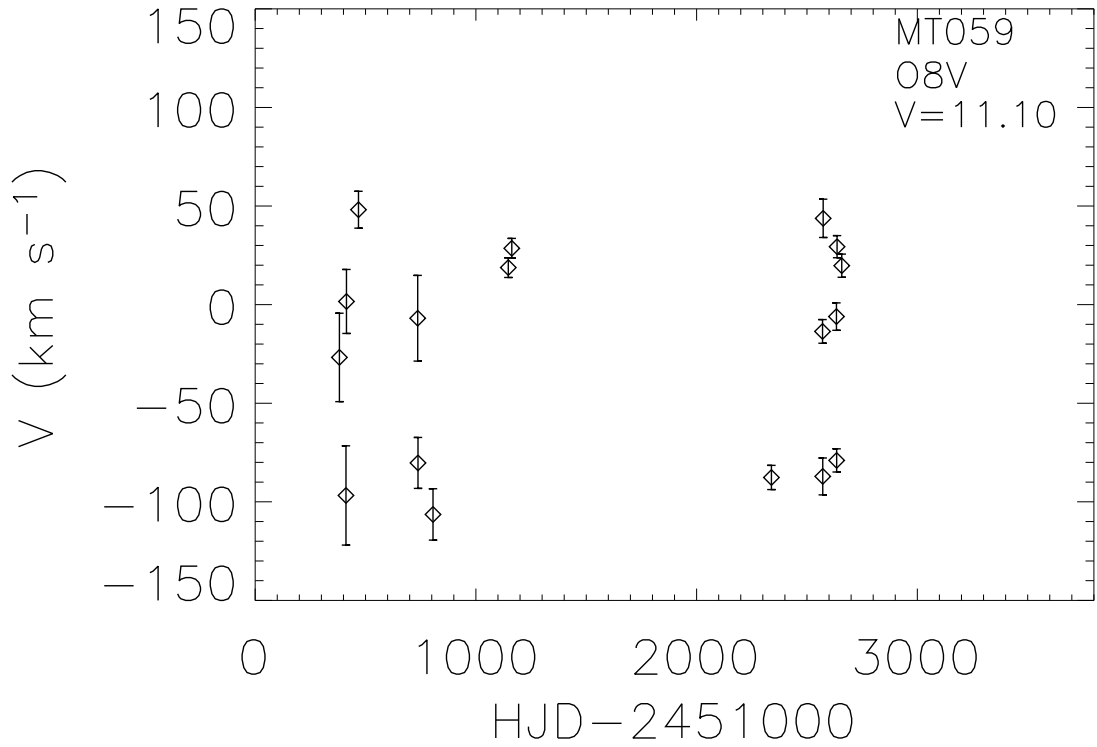}
\caption{Relative radial velocity variation for MT059, one of the more
prominent velocity-variable systems. MT059 is among the \varsystems\ stars with
$P(\chi^2,\nu)\leq0.01$.  \label{59var}}
\end{figure}

\begin{figure}
\epsscale{1.0}
\plotone{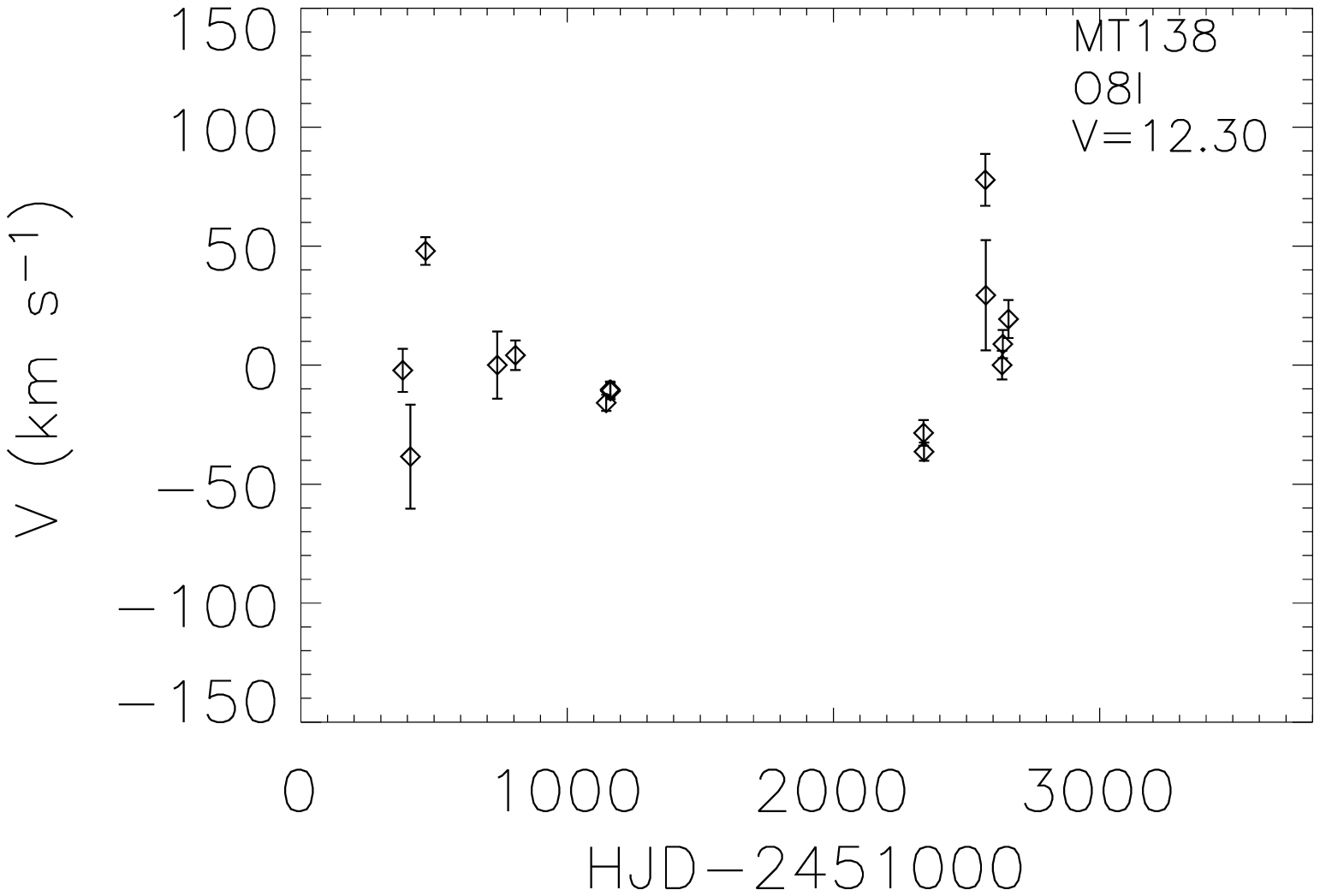}
\caption{Relative radial velocity variation for MT138, one of the more
prominent  velocity-variable systems.  \label{138var}}
\end{figure}

\begin{figure}
\epsscale{1.0}
\plotone{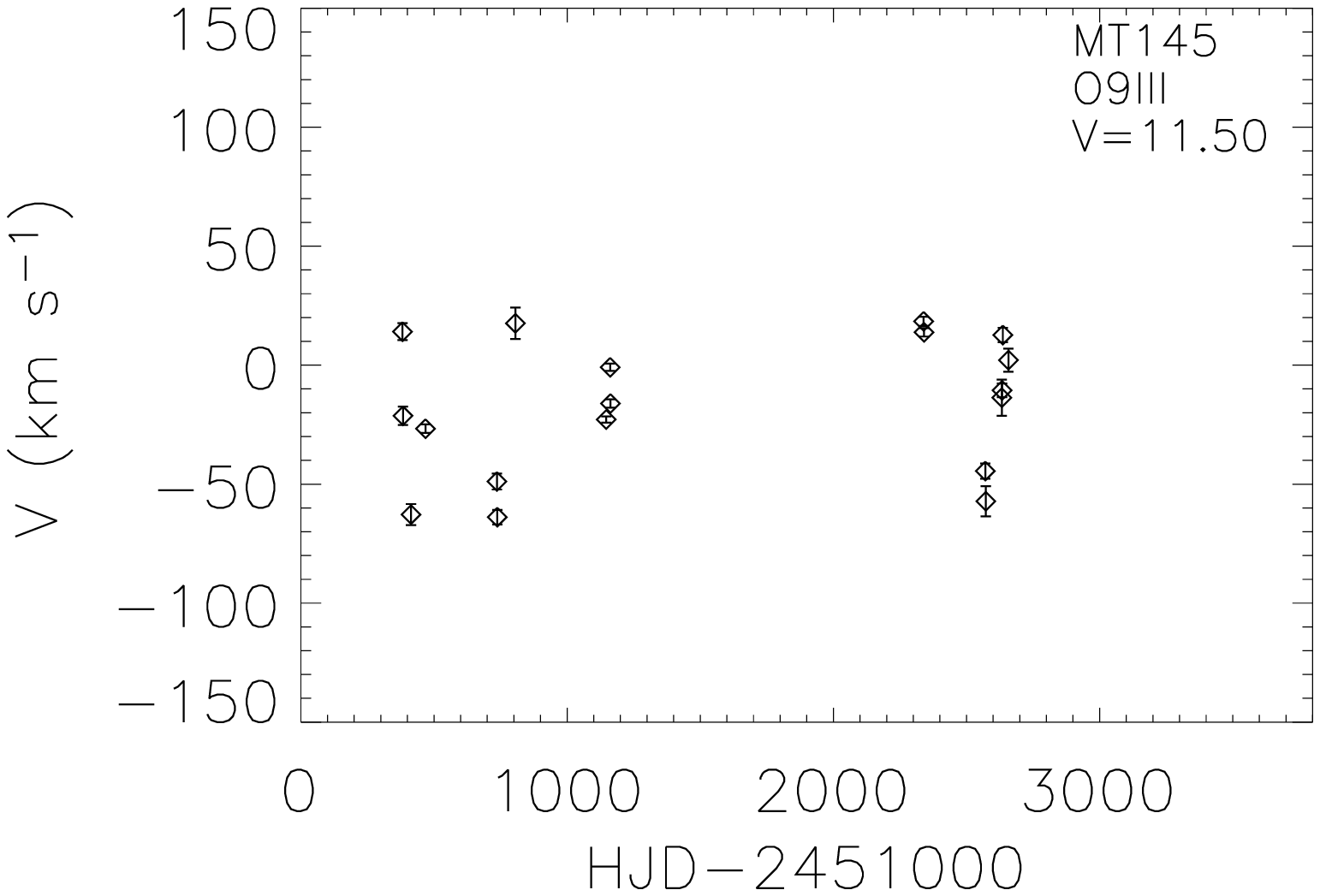}
\caption{Relative radial velocity variation for MT145, one of the more
prominent velocity-variable systems.  \label{145var}}
\end{figure}

\begin{figure}
\epsscale{1.0}
\plotone{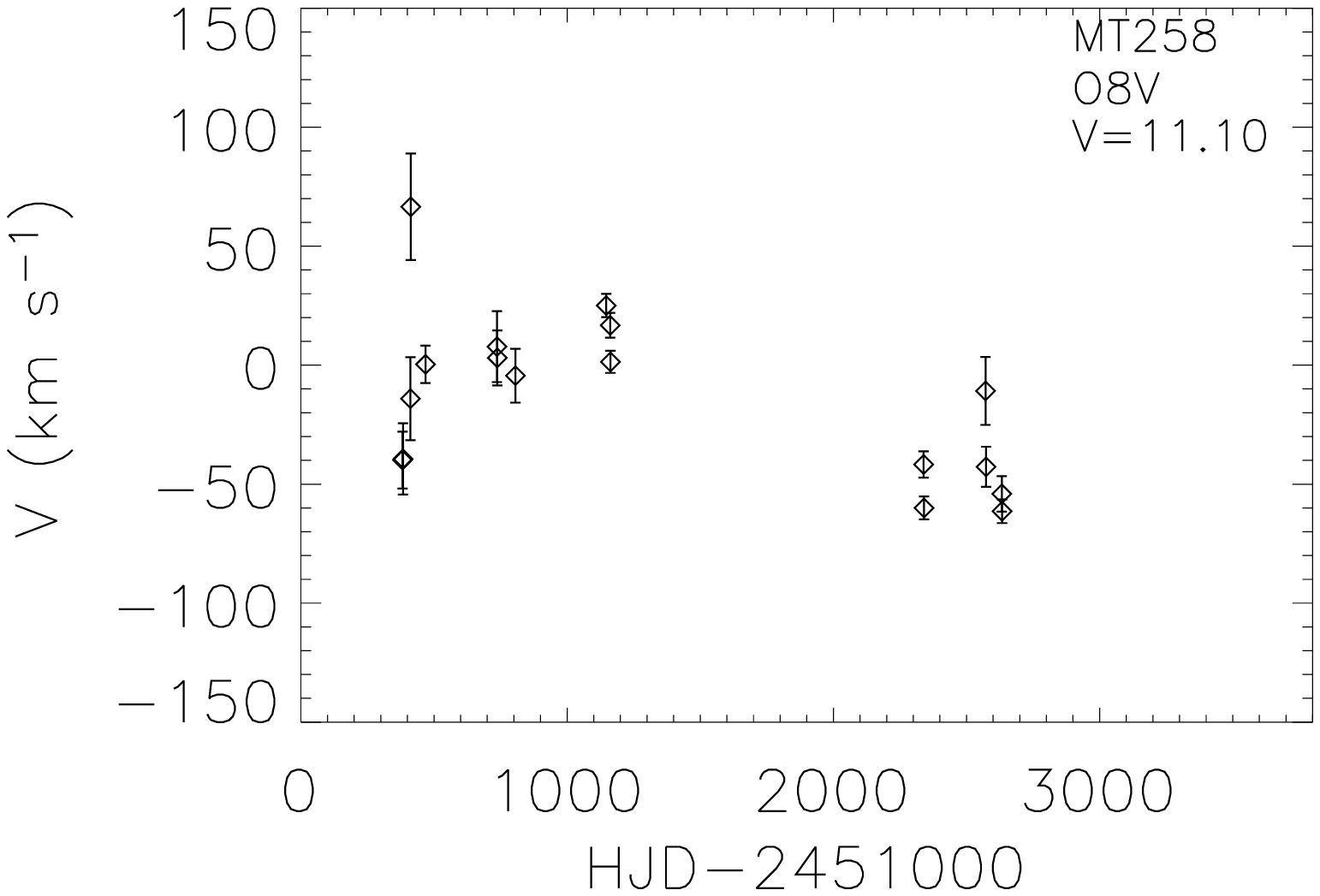}
\caption{Relative radial velocity variation for MT258, one of the more
prominent velocity-variable systems.  \label{258var}}
\end{figure}

\begin{figure}
\epsscale{1.0}
\plotone{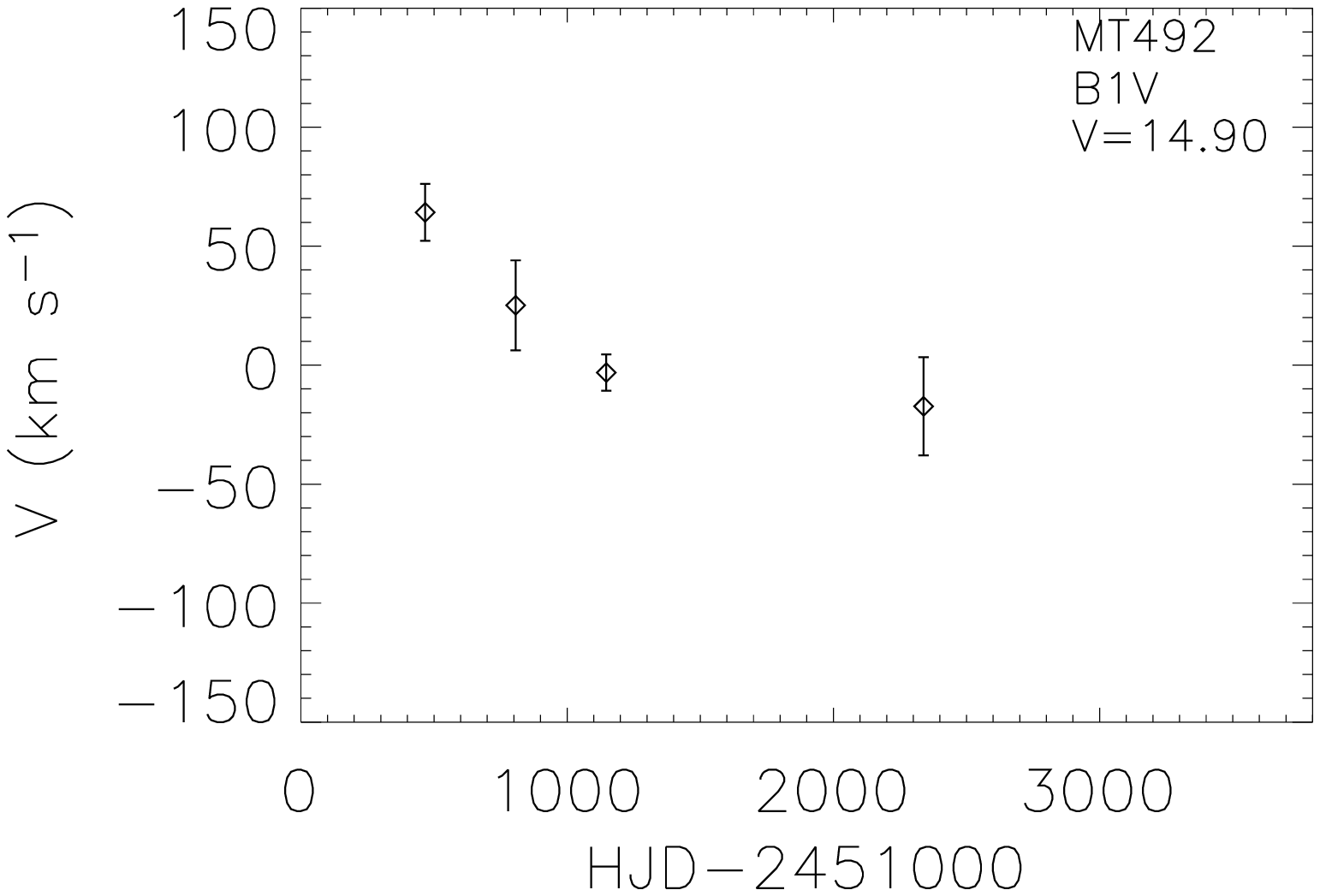}
\caption{Relative radial velocity variation for MT492, one of the more
prominent velocity-variable systems.  \label{492var}}
\end{figure}

\begin{figure}
\epsscale{1.0}
\plotone{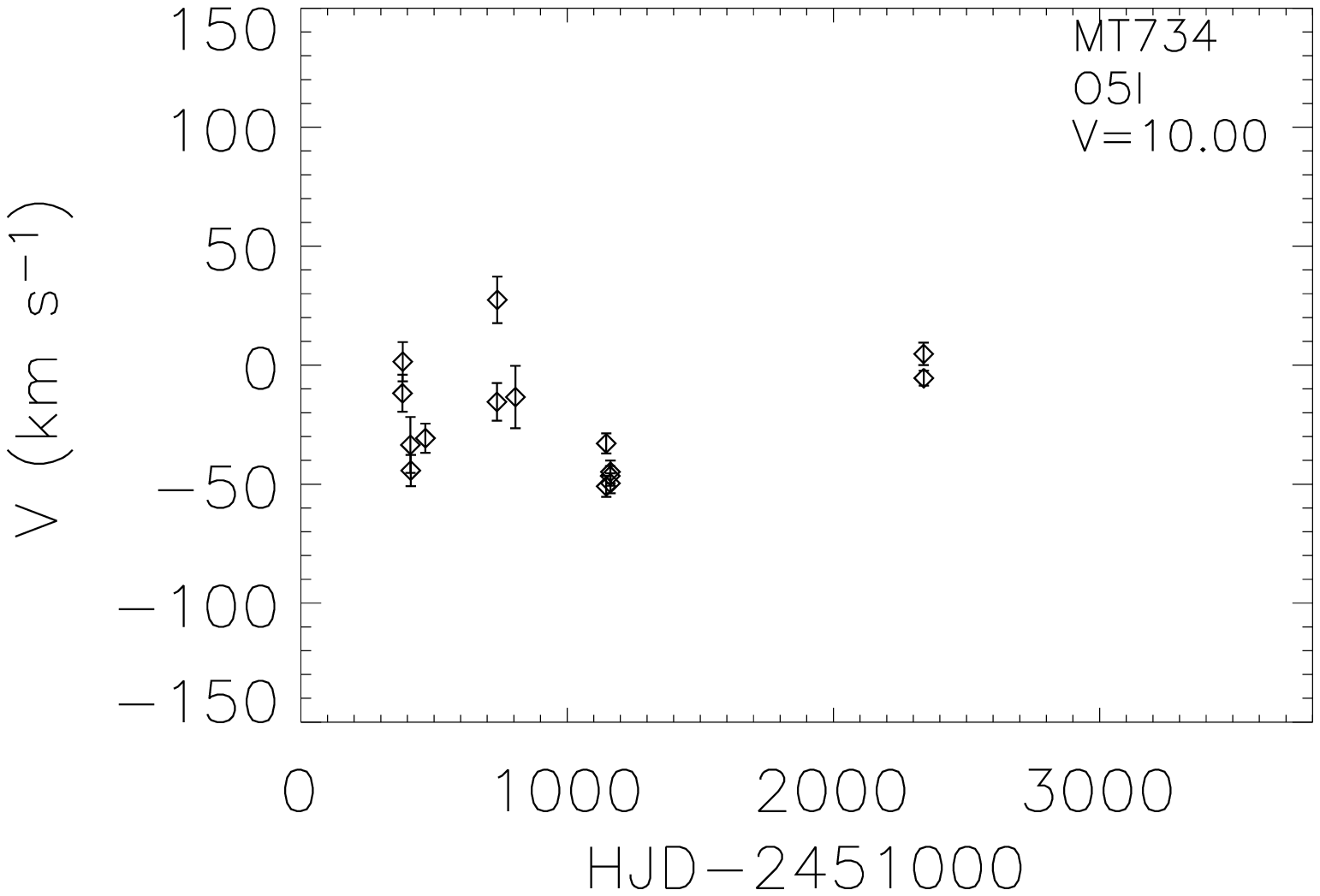}
\caption{Relative radial velocity variation for MT734, one of the more
prominent velocity-variable systems.  \label{734var}}
\end{figure}

Figures~\ref{59var} through \ref{734var} show examples of some of the
more prominent large-amplitude systems.  In Figure~\ref{59var}, we
present the velocity curve of MT059.  This star exhibits large
velocity variations of $\sim$100 \kms\ with an average uncertainty of
$\sim$11 \kms\ on a timescale of $\sim3$ days.  Although our present
data are not sufficiently time-sampled to uniquely determine the
orbital parameters, these rough values imply a secondary mass of
$\geq7$ $M\sun$ and an orbital separation of $\sim0.12$ AU assuming an
inclination near $i=90^o$ and a circular orbit.  Figure~\ref{138var}
shows MT138, an O8I star with an amplitude of nearly 85 \kms, an
average uncertainty of $\sim$9 \kms, and variability on a several day
timescale.  These rough values imply a companion mass of $\geq8$
$M\sun$ and an orbital separation of $\sim0.17$ AU assuming an
inclination near $i=90^o$.  Figure~\ref{145var} shows MT145, another
evolved star (O9III). It has an amplitude of nearly 50 \kms\ and an
average uncertainty of $\sim$4 \kms\ with an apparent period of
several days.  The implied companion parameters are $M\geq 5$ $M\sun$
and $a\simeq0.15$ AU.  Another O8V star (MT258) is presented in
Figure~\ref{258var}. It displays an amplitude of at least 45 \kms\ and
an average uncertainty of $\sim$10 \kms\ with a period of several
days.  The implied companion parameters are $M\sim3$ $M\sun$ and
$a\simeq0.12$ AU.  MT492, a B1V in Figure~\ref{492var} is
representative of early B stars which have relatively few observations
but still strong evidence for velocity variability.  It has an
amplitude of at least 50 \kms\ and an average uncertainty of $\sim$15
\kms.  MT734, shown in Figure~\ref{734var} is one of the brightest and
most massive stars in the survey (O5I). We estimate an amplitude of
$\sim$40 \kms, an average uncertainty of $\sim$7 \kms, and variations
consistent with a period on the order of days.  The implied companion
parameters are $M\geq4$ $M\sun$ and $a\simeq0.15$ AU. At this time we
can only make rough estimates of the orbital parameters for these few
objects with the largest velocity amplitudes. Our present data,
however, do not allow us to rule out short period aliases or 
non-periodic variations.

\subsection{Double-Lined Spectroscopic Binaries}

The cross-correlation analyses reveal \dsbtwos\ systems with
double-lined spectra and \sbtwos\ additional systems with possible
double-lined signatures (\sb2andsb1\ of these are also designated
SB1).  These two groups are designated in column 10 of
Table~\ref{veltable.tab} as SB2 and SB2?, respectively.  The latter
group contains objects with variable cross-correlation widths,
indicating possible multiplicity where the stellar features are
blended at the resolution of the spectra.  The former group contains
objects where the velocity separation of the features is sufficient to
allow the recognition of distinct spectral components at one or more
epochs. Two of these systems are previously identified as SB2 in the
literature: MT465 \citep{Debeck2004} and MT696 \citep{Rios2004}.

MT252 shows two velocity components in the \ion{He}{1} and \ion{Mg}{1}
lines and variable width and asymmetry in the Hydrogen lines during 
two epochs (2001 September 8~--~9). The 2001 September 8
spectrum displays a maximum velocity amplitude of 101 \kms
(\ion{He}{1}) for both components, relative to the mean systemic
velocity. This implies a mass ratio near unity.  The ratio of
luminosities of the two components is also near unity.  Given the
B1.5III spectral classification, this suggests the secondary is
probably another B1.5III or possibly an O9V.

MT465 (Cyg OB2 No.8a) is reported to be a binary consisting of an O6
and an O5.5 star \citep{Debeck2004}. Despite the obvious double-lined
spectra, 23-day period, and 200 \kms\ velocity separation reported in
that work, the components are blended in all of our spectra. We
therefore report this star as an SB1 and include an SB2 designation
only because of the  \citet{Debeck2004} study.

MT696, an 09.5V, is reported to be a double system containing an
early-B secondary \citep{Rios2004}. At least four of our observations
of this star displayed dual velocity components in multiple lines. The
ratio of velocity amplitudes is $\sim0.7$ with a maximum
velocity separation of 540 \kms\ on 2005 September 22. With
a mass 70\% of the O9.5V primary, the deduced secondary spectral type
(B1/B2V) agrees with the findings of \citet{Rios2004}.

MT720, classified as an O9.5V, shows multiple velocity components,
indicating that it is a probable triple system. The 2004 November 30
spectrum displays two velocity features with a separation of 350 \kms\
for the \ion{He}{1} lines.  The 2001 August 24 spectrum also shows
two velocity components, but there is an additional asymmetry on the
redshifted side of the H and He features. The 2001 September 9
spectrum clearly shows at least three velocity components in multiple
lines. Luminosities amongst the primary, secondary, and tertiary are
near unity and suggest comparable masses indicative of a system with
three late-type O stars or B0 stars.

MT771, an O7V, shows \ion{He}{1} double-lined components in three
epochs (2001 September 9 and 2004 November 28--29), with a maximum
velocity separation of 275 \kms. The ratio of velocity
amplitudes is greater than 0.8 as measured about the systemic
velocity calculated in the 2004 November 30 spectrum. The ratio of
luminosities is $\sim$0.75, suggesting that the companion is a late O
star.

\subsection{Velocity Dispersion of Cyg OB2}

\begin{figure}
\epsscale{1.0}
\plotone{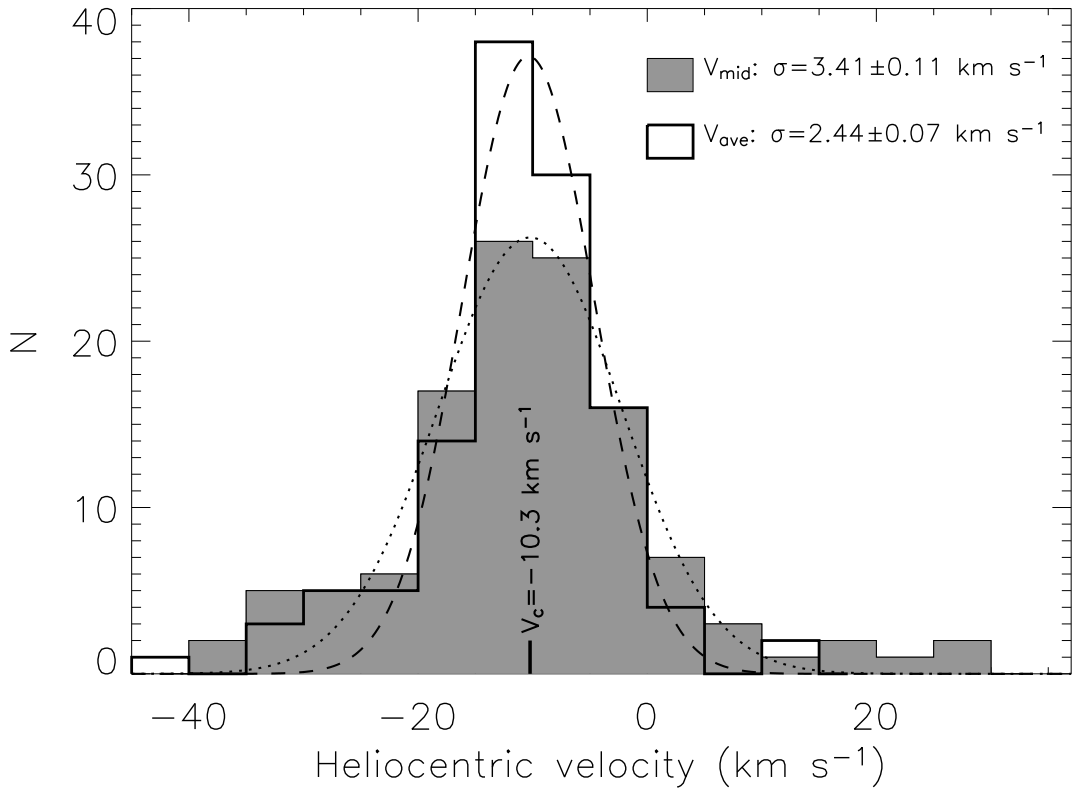}
\caption{Mean systemic velocity distribution for Cygnus OB2 stars
listed in Table~\ref{veltable.tab}.  The unshaded histogram represents
the $V_{avg}$ velocities in column 4.  The shaded histogram represents
the $V_{mid}$ velocities from column 5. Gaussian fits to each
histogram yield a mean systemic velocity of $\sim$-10 \kms, and
dispersions of $\sigma=2.44 \pm 0.07$ \kms\ for $V_{avg}$ and
$\sigma=3.41 \pm 0.11$ for $V_{mid}$. \label{syst}}
\end{figure}

Figure~\ref{syst} provides histograms of systemic velocities for our
two methods of velocity measurement.  The shaded histogram represents
the $V_{mid}$ velocities in column 5 of Table~\ref{veltable.tab}.  The
unshaded histogram represents the $V_{avg}$ velocities from column 4
of Table~\ref{veltable.tab}. A Gaussian fit to the data yields a FWHM
of $8.01 \pm 0.26$ \kms\ for the shaded histogram and $5.7 \pm 0.17$
\kms\ for the unshaded histogram. This translates to a one dimensional
radial velocity dispersion of $\sigma_{V}=3.41 \pm 0.11$ \kms\ and
$\sigma_{V}=2.44 \pm 0.07$ \kms\ for $V_{mid}$ and $V_{avg}$
respectively. Both fits show that the mean systemic velocity of Cyg
OB2 stars is $\overline{V}_{hel}=-10.3 \pm 0.3$ \kms.  Because the
$V_{mid}$ values have more outliers at large relative velocities, this
measurement is best regarded as an upper limit to the true velocity
dispersion.  By comparison, typical open cluster velocity dispersions
lie in the range 1~--~2.5 \kms\ and may depend on initial conditions
\citep{bonnell95}. The velocity dispersions of other open clusters,
such as the Orion nebula cluster \citep[2 \kms]{jones88}, Perseus OB2
\citep[1~--~3 \kms]{steenbrugge2003}, or Scorpius OB2 \citep[1.0 --
1.5 \kms]{debruijne99}, are consistent with our measurement of
$V_{ave}=2.44\pm0.07$ \kms. We also examined systemic velocity as a
function of position within the cluster and found no evidence of large
scale velocity patterns that might indicate the presence of kinematic
subgroups.

\subsection{Runaways in Cyg OB2}

Nearly 20\% of all O stars are runaways, with the other 70~--~80\%
locked in open clusters or OB associations \citep{kaper2005,gies87}.
The origin of runaway stars is still uncertain, but the two leading
theories, asymmetric supernovae ejection and dynamical interaction
require the existence of a companion \citep{blaauw93,blaauw61}.  With
more than 30 O stars listed in Table~\ref{veltable.tab}, and a binary
fraction of at least 50\% among the early type stars
\citep{merm2001,gies87}, one might expect to see 2~--~3 OB runaways in
Cyg OB2. Runaways are usually defined as having space velocities
exceeding 30~--~40 \kms\ \citep{blaauw61} relative to their parent
cluster or mean local Galactic rotation velocity.  Even at minimal
runaway speeds, the Association crossing time is $\leq 0.5$ Myr for a
diameter of $\sim30$ pc.  Thus, any runaways with velocities which are
primarily tangential to the line of sight would travel well beyond the
canonical boundaries of Cyg OB2 within its lifetime. \citet{comeron98}
speculates that some of the OB stars in the Cygnus region that have
high proper motions may have been ejected from Cyg OB2.  Most Cyg OB2
members lack proper motion measurements, and our data are sensitive
only to motions in the radial direction.  Therefore, we would expect,
at most, 1~--~2 runaway stars with large radial velocity components
detectable in this survey.

Figure~\ref{syst} shows that no stars have radial velocities larger
than 35 \kms\ relative to the mean systemic velocity of the
Association.  The most notable outliers, which have relative
velocities of 20~--~30 \kms, are also those with the fewest
measurements and, therefore, are the most uncertain.  We detect no strong
candidates for runaway stars.  Therefore, we conclude that there is
little or no evidence for OB runaways in the radial direction and
likely very few runaways within a $\sim30$ \arcmin\ radius of the
cluster center.

\section{Conclusions}

We conducted a radial velocity survey of the Cygnus OB2 Association
over a 6-year time interval to search for MCBs using spectroscopic
data from the Keck, Lick, WIYN, and WIRO Observatories.  We obtained
\numspectra\ spectra to measure radial velocities and radial velocity
variations on \numstars\ OB stars. There were \newOBstars\ identified
as new early types. The calculated mean distance modulus for Cyg OB2
stars is $\sim$11.3 mag, which is in good agreement with previous
estimates. Of the 146 total OB stars, analysis of the 143 provisional
members yielded an IMF slope of $\Gamma = -2.2 \pm 0.1$. There were a
number of minor spectral classification differences, including Cygnus
OB2 No.~12 and No.~21. No.~12 showed evidence of a B3Iae spectral
classification in at least one epoch and a temperature class variation
(B3~--~B8) over one year.  We utilized two methods for determining
velocity variations, including Gaussian profile fitting and cross
correlation techniques. Both methods yielded similar results, where
\varsystems\ stars had a probability $P(\chi^2,\nu)\leq0.01$ and
\pvarsystems\ stars had a probability of $0.01 < P(\chi^2,\nu)\leq
0.04$. In addition, we detected \dsbtwos\ SB2 systems and \sbtwos\
possible SB2 systems (3 of which were also designated SB1). This
translates to a lower limit on the massive binary frequency of
\varpercent (36 out of 120 stars) to \pvarpercent (50 out of 120 stars). 
The calculated velocity dispersion for Cygnus OB2 is $2.44 \pm 0.07$
\kms, which is typical of open clusters, and despite the richness of
the association and the number of stars surveyed, we detected no
obvious OB runaways.

\acknowledgments

We thank the time allocation committees of the Lick, Keck, WIYN, and
WIRO Observatories for granting us observing time and making this
project possible. This paper has been greatly improved by the detailed 
review and suggestions from an anonymous referee.  
We are grateful for support from the National
Science Foundation through the Research Experiences for Undergraduates
(REU) program grant AST-0353760 and through grant AST-0307778, and the
support of the Wyoming NASA Space Grant Consortium.  The work of
C.F. was in part under the auspices of the U.S.\ Dept.\ of Energy, and
supported by its contract W-7405-ENG-36 to Los Alamos National
Laboratory and by National Science Foundation under Grant
No. PHY99-07949.

{\it Facilities:} \facility{WIRO ()}, \facility{WIYN ()}, \facility{Shane ()}, 
\facility{Keck:I ()}

{}

\clearpage

\begin{deluxetable}{llll}
\tabletypesize{\scriptsize}
\tablecaption{Observing Information \label{obs.tab}}
\tablewidth{0pt}
\tablehead{
\colhead{Date} &
\colhead{Observatory/Instrument} &
\colhead{Spectral coverage} &  
\colhead{HJD coverage}}   
\startdata
1999 July 4~--~5         & Keck/HIRES      & 3890~--~6270~\AA\ in 35 orders  & 2451363~--~2451364 \\
1999 July 21~--~23       & Lick/Hamilton   & 3650~--~7675~\AA\ in 81 orders  & 2451381~--~2451383 \\
1999 August 21~--~23     & Lick/Hamilton   & 3650~--~7675~\AA\ in 81 orders  & 2451411~--~2451413 \\
1999 October 14~--~15    & Keck/HIRES      & 3700~--~5250~\AA\ in 29 orders  & 2451466~--~2451467 \\
2000 July 10~--~11       & Lick/Hamilton   & 3650~--~7675~\AA\ in 81 orders  & 2451736~--~2451737 \\
2000 September 18~--~19  & Keck/HIRES      & 3700~--~5250~\AA\ in 29 orders  & 2451805~--~2451806 \\
2001 August 24           & WIYN/Hydra      & 3800~--~4490~\AA\ in order 2    & 2452146            \\
2001 September 8~--~9    & WIYN/Hydra      & 3800~--~4490~\AA\ in order 2    & 2452161~--~2452162 \\
2004 November 28~--~30   & WIYN/Hydra      & 3800~--~4490~\AA\ in order 2    & 2453338~--~2453340 \\
2005 July 18~--~21       & WIRO/WIRO-Spec  & 3800~--~4490~\AA\ in order 1    & 2453570~--~2453573 \\
2005 July 18~--~20,22    & WIRO/WIRO-Spec  & 3800~--~4490~\AA\ in order 1    & 2453632~--~2453635 \\
2005 October 13          & WIRO/Longslit   & 4050~--~6050~\AA\ in order 2    & 2453657
\enddata
\end{deluxetable}

\clearpage
\pagestyle{empty}
\begin{deluxetable}{
l@{\hspace{0.5em}}
l@{\hspace{0.1em}}
l@{\hspace{0.1em}}
l@{\hspace{1.0em}}
l@{\hspace{0.1em}}
c@{\hspace{0.1em}}
c@{\hspace{0.1em}}
c@{\hspace{0.1em}}
c@{\hspace{0.1em}}
c@{\hspace{0.1em}}
c@{\hspace{0.1em}}
c@{\hspace{0.1em}}
c@{\hspace{0.3em}}
c@{\hspace{1.0em}}
c@{\hspace{1.0em}}
c}
\tabletypesize{\scriptsize} 
\tablecaption{Early Type Stars in the Direction of Cyg OB2 
\label{bigtable.tab}} 
\tablewidth{0pt} 
\rotate
\tablehead{ 
\colhead{Star} & 
\colhead{Lit S.C.} &
\colhead{Our S.C.} & 
\colhead{RA} &
\colhead{Dec} & 
\colhead{$\mathrm{A}_{V}$} & 
\colhead{V} &
\colhead{$\mathrm{M}_{V}$} & 
\colhead{(\bv)} & 
\colhead{$\mathrm{(\bv)}_0$} &
\colhead{$\mathrm{M}_{\mathrm{PD}}$} &
\colhead{$\mathrm{M}_0$} &  
\colhead{DM} &
\colhead{Notes} &
\colhead{Ph. Activ.} &
\colhead{Ref.} \\
\colhead{(MT)} & 
\colhead{} &
\colhead{} &
\colhead{(J2000)} & 
\colhead{(J2000)} & 
\colhead{} & 
\colhead{} &
\colhead{} & 
\colhead{} & 
\colhead{} & 
\colhead{($\mathrm{M}_\odot$)} &
\colhead{($\mathrm{M}_\odot$)} & 
\colhead{} & 
\colhead{} &
\colhead{} &
\colhead{} \\
\colhead{(1)} & 
\colhead{(2)} &
\colhead{(3)} &
\colhead{(4)} & 
\colhead{(5)} & 
\colhead{(6)} & 
\colhead{(7)} &
\colhead{(8)} & 
\colhead{(9)} & 
\colhead{(10)} & 
\colhead{(11)} &
\colhead{(12)} & 
\colhead{(13)} & 
\colhead{(14)} &
\colhead{(15)} &
\colhead{(16)}}

\startdata 
5 & O6V & O6V              & 20 30 39.87 & +41 36 50.9 & 5.8 & 12.93 & -4.92    & 1.64 & -0.30 & 31.7 & 31.7 & 12.0 &     &     &    \\ 
20 & \nodata & B0V         & 20 30 51.12 & +41 20 21.6 & 7.3 & 14.48 & -3.80    & 2.18 & -0.26 & 17.5 & 17.5 & 10.9 & n   &     &    \\ 
21 & B(2?) & B2II          & 20 30 50.75 & +41 35 06.0 & 4.4 & 13.74 & -4.80    & 1.30 & -0.18 & 18.9 & 20.0 & 14.1 &     &     &    \\ 
42 & \nodata & B2V       & 20 30 59.43 & +41 35 59.6 & 4.4 & 13.73 & -2.50    & 1.27 & -0.21 & 10.0 & 10.0 & 11.7 & n   &     &    \\ 
59 & O8.5V & O8V           & 20 31 10.57 & +41 31 53.0 & 5.3 & 11.18 & -4.34    & 1.47 & -0.28 & 22.0 & 22.0 & 10.2 &     &     &    \\
70 & O9V & O9V             & 20 31 18.31 & +41 21 21.7 & 7.1 & 12.99 & -4.05    & 2.10 & -0.28 & 18.0 & 18.0 & 10.0 &     &     &    \\ 
83 & B1I & B1I             & 20 31 22.03 & +41 31 28.0 & 4.1 & 10.64 & -6.50    & 1.18 & -0.19 & 24.0 & 26.0 & 13.0 &     & Susp  & 1   \\ 
97 & \nodata & B2V         & 20 31 30.49 & +41 37 15.6 & 4.3 & 14.56 & -2.50    & 1.22 & -0.21 & 10.0 & 10.0 & 12.7 & n   &     &    \\ 
103 & \nodata & B1V        & 20 31 33.38 & +41 22 49.0 & 6.7 & 13.81 & -3.20    & 2.00 & -0.23 & 13.8 & 13.8 & 10.3 & n   &     &    \\ 
106 & \nodata & B3V        & 20 31 33.59 & +41 36 04.3 & 4.4 & 14.60 & -1.90    & 1.29 & -0.18 &  7.6 &  7.6 & 12.0 & n   &     &    \\ 
108 & \nodata & B3IV       & 20 31 34.12 & +41 31 08.0 & 4.3 & 14.88 & -2.50    & 1.26 & -0.16 &  7.6 &  7.6 & 13.1 & n   &     &    \\ 
129 & \nodata & B3V        & 20 31 41.60 & +41 28 20.9 & 4.6 & 14.40 & -1.90    & 1.34 & -0.18 &  7.6 &  7.6 & 11.7 & n   &     &    \\ 
138 & O8.5I & O8I          & 20 31 45.39 & +41 18 26.8 & 6.9 & 12.26 & -6.30    & 1.99 & -0.30 & 36.8 & 39.8 & 11.6 &     &     &    \\ 
145 & O9.5V & O9III        & 20 31 49.65 & +41 28 26.8 & 4.1 & 11.52 & -5.25    & 1.11 & -0.26 & 23.1 & 24.6 & 12.6 &     &     &    \\ 
164 & \nodata & B3V        & 20 31 55.28 & +41 35 27.8 & 4.3 & 15.07 & -1.90    & 1.25 & -0.18 &  7.6 &  7.6 & 12.6 & n   &     &    \\ 
169 & B1.5V & B2V          & 20 31 56.27 & +41 33 05.3 & 4.3 & 13.90 & -2.50    & 1.21 & -0.21 & 10.0 & 10.0 & 12.1 &     &     &    \\ 
170 & \nodata & B(5?)V     & 20 31 56.23 & +41 35 12.3 & 3.6 & 15.21 & -1.35    & 1.05 & -0.15 &  5.9 &  5.9 & 11.6 & n   &     &    \\ 
174 & B2V & B2IV           & 20 31 56.90 & +41 31 48.0 & 4.3 & 12.55 & -3.10    & 1.21 & -0.21 & 10.0 & 10.0 & 11.3 &     &     &    \\ 
179 & \nodata & B3V        & 20 31 59.82 & +41 37 14.3 & 4.9 & 14.00 & -1.90    & 1.46 & -0.18 &  7.6 &  7.6 & 10.9 & n   &     &    \\ 
186 & \nodata & B2Ve       & 20 32 03.01 & +41 32 30.7 & 4.2 & 14.14 & -2.50    & 1.20 & -0.21 & 10.0 & 10.0 & 12.4 & n   &     &    \\ 
187 & B1V & B1V            & 20 32 03.74 & +41 25 10.9 & 5.3 & 13.24 & -3.20    & 1.52 & -0.23 & 13.8 & 13.8 & 11.1 &     &     &    \\ 
189 & \nodata & B(6?)V     & 20 32 04.57 & +41 27 48.6 & 3.9 & 14.77 & -1.15    & 1.17 & -0.14 &  5.4 &  5.4 & 10.8 & n   &     &    \\ 
191 & \nodata & B3IV       & 20 32 04.74 & +41 28 44.5 & 3.6 & 13.81 & -2.50    & 1.04 & -0.16 &  7.6 &  7.6 & 12.7 & n   &     &    \\ 
196 & \nodata & B6V        & 20 32 05.59 & +41 27 49.6 & 4.4 & 14.81 & -1.15    & 1.31 & -0.14 &  5.4 &  5.4 & 10.4 & n   &     &    \\ 
200 & \nodata & B3V        & 20 32 06.85 & +41 17 56.8 & 6.5 & 13.94 & -1.90    & 1.98 & -0.18 &  7.6 &  7.6 &  9.4 & n   &     &    \\ 
202 & \nodata & B2V        & 20 32 07.95 & +41 22 00.3 & 4.7 & 14.40 & -2.50    & 1.35 & -0.21 & 10.0 & 10.0 & 12.2 & n   &     &    \\ 
213 & B0V & B0V            & 20 32 13.07 & +41 27 24.9 & 4.2 & 11.95 & -3.80    & 1.13 & -0.26 & 17.5 & 17.5 & 11.5 &     &     &    \\ 
215 & B1V & B2V            & 20 32 13.48 & +41 27 31.0 & 3.5 & 12.97 & -2.50    & 0.96 & -0.21 & 10.0 & 10.0 & 11.9 &     &     &    \\ 
216 & \nodata & B1.5V      & 20 32 13.75 & +41 27 42.0 & 4.2 & 13.02 & -2.80    & 1.18 & -0.22 & 11.9 & 11.9 & 11.6 & n   &     &    \\ 
217 & O7IIIf & O7IIIf      & 20 32 13.77 & +41 27 12.7 & 4.4 & 10.23 & -5.54    & 1.19 & -0.29 & 31.2 & 33.0 & 11.3 &     &     &    \\ 
220 & \nodata & B1V        & 20 32 14.56 & +41 22 33.7 & 5.3 & 14.34 & -3.20    & 1.52 & -0.23 & 13.8 & 13.8 & 12.2 & n   &     &    \\ 
221 & \nodata & B2V        & 20 32 14.63 & +41 27 40.3 & 4.5 & 13.62 & -2.50    & 1.30 & -0.21 & 10.0 & 10.0 & 11.5 & n   &     &    \\
222 & \nodata & B3V        & 20 32 15.03 & +41 19 30.8 & 5.0 & 14.80 & -1.90    & 1.47 & -0.18 &  7.6 &  7.6 & 11.7 & n   &     &    \\ 
227 & O9V & O9V            & 20 32 16.53 & +41 25 36.4 & 4.6 & 11.47 & -4.05    & 1.24 & -0.28 & 18.0 & 18.0 & 10.9 &     &     &    \\ 
234 & \nodata & B2V        & 20 32 19.66 & +41 20 39.7 & 4.5 & 13.25 & -2.50    & 1.28 & -0.21 & 10.0 & 10.0 & 11.2 & n   &     &    \\ 
238 & \nodata & B1V        & 20 32 21.35 & +41 18 35.5 & 6.1 & 14.91 & -3.20    & 1.80 & -0.23 & 13.8 & 13.8 & 12.0 & n   &     &    \\ 
239 & \nodata & B4V        & 20 32 21.76 & +41 34 24.6 & 3.9 & 14.33 & -1.65    & 1.14 & -0.16 &  6.4 &  6.4 & 10.4 & n   &     &    \\ 
241 & \nodata & B2V        & 20 32 22.15 & +41 27 41.7 & 4.0 & 13.41 & -2.50    & 1.13 & -0.21 & 10.0 & 10.0 & 11.8 & n   &     &    \\ 
248 & \nodata & B2V        & 20 32 25.50 & +41 24 51.8 & 4.6 & 13.36 & -2.50 	& 1.31 & -0.21 & 10.0 & 10.0 & 11.2 &     &     &    \\ 
250 & B1V & B2III          & 20 32 26.10 & +41 29 39.0 & 3.8 & 12.88 & -3.70 	& 1.06 & -0.19 & 14.8 & 15.0 & 12.8 &     &     &    \\ 
252 & \nodata & B1.5III    & 20 32 26.50 & +41 19 13.7 & 4.9 & 14.15 & -3.90 	& 1.42 & -0.20 & 16.1 & 16.3 & 13.1 & n   &     &    \\ 
255 & \nodata & B2III      & 20 32 27.26 & +41 21 56.2 & 4.8 & 14.71 & -3.70 	& 1.42 & -0.19 & 14.8 & 15.0 & 13.5 & n   &     &    \\ 
258 & O8V & O8V            & 20 32 27.67 & +41 26 21.7 & 4.5 & 11.10 & -4.34 	& 1.20 & -0.28 & 22.0 & 22.0 & 10.9 &     &     &    \\ 
259 & B1V & B0Ib           & 20 32 27.76 & +41 28 51.9 & 3.7 & 11.42 & -6.00 	& 1.00 & -0.22 & 25.0 & 27.0 & 13.7 &     &     &    \\ 
264 & \nodata & B2III      & 20 32 30.72 & +41 07 04.1 & 3.5 & 12.63 & -3.70 	& 0.99 & -0.19 & 14.8 & 15.0 & 12.7 & n   &     &    \\ 
268 & \nodata & B2.5V      & 20 32 31.42 & +41 30 51.4 & 4.9 & 14.38 & -2.20 	& 1.43 & -0.19 &  8.8 &  8.8 & 11.7 & n   &     &    \\ 
271 & \nodata & B4V        & 20 32 32.34 & +41 22 57.6 & 5.0 & 14.57 & -1.65    & 1.51 & -0.16 &  6.4 &  6.4 &  9.5 & n   &     &    \\ 
273 & \nodata & B(5?)V     & 20 32 32.54 & +41 26 46.7 & 5.2 & 14.91 & -1.35    & 1.57 & -0.15 &  5.9 &  5.9 &  9.7 & n   &     &    \\ 
275 & \nodata & B2V        & 20 32 32.68 & +41 27 04.4 & 3.9 & 13.47 & -2.50 	& 1.10 & -0.21 & 10.0 & 10.0 & 12.0 & n   &     &    \\ 
292 & B1V & B2V            & 20 32 37.03 & +41 23 05.1 & 5.2 & 13.08 & -2.50 	& 1.51 & -0.21 & 10.0 & 10.0 & 10.4 &     &     &    \\ 
295 & \nodata & B2V        & 20 32 37.78 & +41 26 15.3 & 4.3 & 13.71 & -2.50 	& 1.21 & -0.21 & 10.0 & 10.0 & 11.9 & n   &     &    \\ 
298 & \nodata & B3V        & 20 32 38.34 & +41 28 56.6 & 4.3 & 14.43 & -1.90 	& 1.25 & -0.18 &  7.6 &  7.6 & 12.0 & n   &     &    \\ 
299 & O7V & O7V            & 20 32 38.58 & +41 25 13.6 & 4.4 & 10.84 & -4.63 	& 1.19 & -0.29 & 26.5 & 26.5 & 11.0 &     &     &    \\ 
300 & B(1?) & B1V          & 20 32 38.87 & +41 25 20.8 & 4.3 & 13.05 & -3.20 	& 1.21 & -0.23 & 13.8 & 13.8 & 11.9 &     &     &    \\ 
304 & B5Iae & B3Iae        & 20 32 40.88 & +41 14 29.3 & 10. & 11.46 & -6.30 	& 3.35 & -0.13 & 22.0 & 23.5 &  7.4 &     & Susp & 1 \\ 
311 & \nodata & B2V        & 20 32 42.90 & +41 20 16.4 & 4.8 & 13.87 & -2.50 	& 1.39 & -0.21 & 10.0 & 10.0 & 11.5 & n   &     &    \\ 
317 & O8V & O8V            & 20 32 45.45 & +41 25 37.3 & 4.6 & 10.68 & -4.34 	& 1.25 & -0.28 & 22.0 & 22.0 & 10.4 &     &     &    \\ 
322 & \nodata & B2.5V      & 20 32 46.45 & +41 24 22.4 & 4.6 & 14.91 & -2.20 	& 1.33 & -0.19 &  8.8 &  8.8 & 12.5 & n   &     &    \\ 
325 & \nodata & B1.5III    & 20 32 46.74 & +41 26 15.9 & 4.7 & 14.30 & -3.90 	& 1.37 & -0.20 & 16.1 & 16.3 & 13.4 & n   &     &    \\ 
336 & \nodata & B3III      & 20 32 49.67 & +41 25 36.4 & 4.0 & 14.13 & -3.00 	& 1.17 & -0.16 & 12.2 & 12.3 & 13.1 & n   &     &    \\ 
339 & O8.5V & O8V          & 20 32 50.03 & +41 23 44.6 & 4.9 & 11.60 & -4.34 	& 1.35 & -0.28 & 22.0 & 22.0 & 11.0 &     &     &    \\ 
343 & \nodata & B1V        & 20 32 50.69 & +41 15 02.2 & 6.6 & 14.44 & -3.20 	& 1.98 & -0.23 & 13.8 & 13.8 & 11.0 & n   &     &    \\ 
358 & B & B3V              & 20 32 54.35 & +41 15 22.1 & 7.0 & 14.81 & -1.90 	& 2.16 & -0.18 &  7.6 &  7.6 &  9.7 &     &     &    \\ 
365 & \nodata & B1V        & 20 32 56.66 & +41 23 41.0 & 4.5 & 13.81 & -3.20 	& 1.28 & -0.23 & 13.8 & 13.8 & 12.4 & n   &     &    \\ 
372 & \nodata & B0V        & 20 32 58.79 & +41 04 29.9 & 7.3 & 14.97 & -3.80 	& 2.17 & -0.26 & 17.5 & 17.5 & 11.4 & n   &     &    \\
376 & O8V & O8V            & 20 32 59.17 & +41 24 25.7 & 4.9 & 11.91 & -4.34 	& 1.35 & -0.28 & 22.0 & 22.0 & 11.3 &     &     &    \\
378 & B0V & B0V            & 20 32 59.61 & +41 15 14.6 & 7.1 & 13.49 & -3.80 	& 2.10 & -0.26 & 17.5 & 17.5 & 10.2 &     &     &    \\ 
390 & O8V & O8V            & 20 33 02.94 & +41 17 43.3 & 6.8 & 12.95 & -4.34 	& 1.98 & -0.28 & 22.0 & 22.0 & 10.4 &     & Irr & 2  \\ 
395 & B1.5V & B1V          & 20 33 04.42 & +41 17 08.9 & 6.0 & 14.14 & -3.20 	& 1.75 & -0.23 & 13.8 & 13.8 & 11.4 & n   &     &    \\ 
400 & B & B1V              & 20 33 05.22 & +41 17 51.6 & 5.6 & 14.15 & -3.20 	& 1.62 & -0.23 & 13.8 & 13.8 & 11.8 &     &     &    \\ 
403 & B2V & B1V            & 20 33 05.55 & +41 43 37.2 & 5.2 & 12.94 & -3.20 	& 1.49 & -0.23 & 13.8 & 13.8 & 10.9 &     &     &    \\ 
409 & \nodata & B0.5V      & 20 33 06.62 & +41 21 13.3 & 6.0 & 14.21 & -3.80 	& 1.76 & -0.24 & 15.6 & 15.6 & 12.0 & n   &     &    \\ 
417 & O4III & O4III      & 20 33 08.78 & +41 13 18.1 & 7.1 & 11.55 & -5.98 	& 2.04 & -0.31 & 48.8 & 51.0 & 10.4 &     &     &    \\ 
420 & \nodata & O9V        & 20 33 09.41 & +41 12 58.2 & 6.8 & 12.84 & -4.05 	& 1.97 & -0.28 & 18.0 & 18.0 & 10.1 & n,b &     &    \\ 
421 & O9V & O9V            & 20 33 09.58 & +41 13 00.6 & 6.7 & 12.86 & -4.05 	& 1.96 & -0.28 & 18.0 & 18.0 & 10.1 & b   & EA, $P = 4.161d$ & 2 \\
425 & B0V & B0V            & 20 33 10.10 & +41 13 10.1 & 6.6 & 13.62 & -3.80 	& 1.94 & -0.26 & 17.5 & 17.5 & 10.8 &     &     &    \\ 
426 & B0V & BOV            & 20 33 10.34 & +41 13 06.4 & 6.6 & 14.05 & -3.80 	& 1.95 & -0.26 & 17.5 & 17.5 & 11.2 &     &     &    \\ 
427 & \nodata & B4II-B4III & 20 33 10.27 & +41 23 44.9 & 4.6 & 14.97 & -4.60    & 1.39 & -0.13 & 15.3 & 15.5 & 10.4 & n   &     &    \\ 
428 & \nodata & B1V        & 20 33 10.46 & +41 20 57.6 & 6.0 & 14.06 & -3.20    & 1.77 & -0.23 & 13.8 & 13.8 & 11.2 & n   &     &    \\ 
429 & B0V & B0V            & 20 33 10.50 & +41 22 22.8 & 5.5 & 12.98 & -3.80    & 1.56 & -0.26 & 17.5 & 17.5 & 11.3 &     & EA; $P=2.9788d:$ & 2   \\
431 & O5If & O5If          & 20 33 10.74 & +41 15 08.0 & 6.4 & 10.96 & -6.33    & 1.81 & -0.32 & 50.9 & 53.0 & 10.9  &     & Unk; $P=1.22/5.6d$    & 2   \\ 
435 & \nodata & B0V        & 20 33 11.02 & +41 10 31.9 & 7.4 & 14.78 & -3.80    & 2.19 & -0.26 & 17.5 & 17.5 & 11.2  & n   &     &    \\ 
441 & \nodata & B2(III?)   & 20 33 11.39 & +41 17 58.9 & 5.1 & 14.38 & -3.70    & 1.52 & -0.19 & 14.8 & 15.0 & 12.9  & n   &     &    \\ 
444 & \nodata & B5V        & 20 33 11.81 & +41 24 05.8 & 4.9 & 14.12 & -1.35    & 1.48 & -0.15 &  5.9 &  5.9 & 9.3 & n   &     &    \\ 
448 & O6V & O6V            & 20 33 13.25 & +41 13 28.6 & 7.4 & 13.61 & -4.92    & 2.15 & -0.30 & 31.7 & 31.7 & 11.1  &     & Unk; $P=3.16d?$  & 2   \\ 
453 & \nodata & B(5?)V     & 20 33 13.37 & +41 26 39.7 & 3.9 & 14.45 & -1.35    & 1.14 & -0.15 &  5.9 &  5.9 & 10.5 & n   &     &    \\
455 & O8V & O8V            & 20 33 13.67 & +41 13 05.7 & 6.3 & 12.92 & -4.34    & 1.81 & -0.28 & 22.0 & 22.0 & 10.9 &     &     &    \\ 
457 & O3If & O3If          & 20 33 14.16 & +41 20 21.5 & 5.4 & 10.55 & -6.35    & 1.45 & -0.34 & 80.0 & 100  & 11.5 &     & Irr & 2   \\ 
459 & \nodata & B(5?)      & 20 33 14.34 & +41 19 33.0 & 5.8 & 14.67 & -1.35    & 1.79 & -0.15 &  5.9 &  5.9 &  8.8 & n   &     &    \\ 
462 & O6.5III & O7III-II   & 20 33 14.84 & +41 18 41.4 & 5.2 & 10.33 & -5.54    & 1.44 & -0.29 & 31.2 & 33.0 & 10.6 &     & Cst & 2   \\ 
465 & O5.5I & O5.5I        & 20 33 15.18 & +41 18 50.1 & 4.9 &  9.06 & -6.33    & 1.30 & -0.32 & 48.3 & 50.3 & 10.5 &     &     &    \\
467 & B1V & B1V            & 20 33 15.37 & +41 29 56.6 & 5.5 & 13.43 & -3.20    & 1.59 & -0.23 & 13.8 & 13.8 & 11.1 &     &     &    \\ 
469 & \nodata & B1III      & 20 33 15.51 & +41 27 32.9 & 4.9 & 13.65 & -4.30    & 1.41 & -0.21 & 17.5 & 18.3 & 13.0 & n   &     &    \\ 
470 & O9V & O9V            & 20 33 15.74 & +41 20 17.2 & 5.2 & 12.50 & -4.05    & 1.46 & -0.28 & 18.0 & 18.0 & 11.3 &     & Cst & 2   \\ 
473 & O8.5V & O8.5V        & 20 33 16.36 & +41 19 01.9 & 5.2 & 12.02 & -4.19    & 1.45 & -0.28 & 19.8 & 19.8 & 11.0 &     & Cst & 2   \\ 
477 & \nodata & B(0?)V     & 20 33 17.40 & +41 12 38.7 & 6.5 & 14.43 & -3.80    & 1.91 & -0.26 & 17.5 & 17.5 & 11.7 & n   &     &    \\ 
480 & O7V & O7V            & 20 33 17.49 & +41 17 09.2 & 5.6 & 11.88 & -4.63    & 1.59 & -0.29 & 26.5 & 26.5 & 10.8 &     & Cst & 2   \\ 
483 & O5I & O5III          & 20 33 18.02 & +41 18 31.0 & 4.6 & 10.19 & -5.84    & 1.24 & -0.30 & 41.5 & 43.0 & 11.4 &     & Cst & 2   \\ 
485 & O8V & O8V            & 20 33 18.08 & +41 21 36.6 & 5.4 & 12.06 & -4.34    & 1.51 & -0.28 & 22.0 & 22.0 & 11.0 &     & Cst & 2   \\ 
488 & Be & B1Ve-B3Ve       & 20 33 18.55 & +41 15 35.4 & 7.7 & 14.88 & -3.20    & 2.36 & -0.21 & 10.0 & 10.0 & 10.3 &     & BE  & 2  \\ 
490 & \nodata & B(0?)      & 20 33 18.56 & +41 24 49.3 & 5.4 & 14.76 & -3.80    & 1.53 & -0.26 & 17.5 & 17.5 & 13.1 & n   &     &    \\
492 & \nodata & B1V        & 20 33 19.16 & +41 17 44.9 & 5.7 & 14.85 & -3.20    & 1.68 & -0.23 & 13.8 & 13.8 & 12.3 & n   &     &    \\ 
493 & \nodata & B5IV       & 20 33 19.26 & +41 24 44.8 & 5.3 & 14.99 & -1.35    & 1.61 & -0.15 &  5.9 &  5.9 &  9.7 & n   &     &    \\ 
507 & O9V & O9V            & 20 33 21.04 & +41 17 40.1 & 5.5 & 12.70 & -4.05    & 1.54 & -0.28 & 18.0 & 18.0 & 11.2 &     & Cst & 2   \\ 
509 & \nodata & B0III-B0IV & 20 33 21.14 & +41 35 52.0 & 6.1 & 14.72 & -5.00    & 1.79 & -0.23 & 20.0 & 20.8 & 13.6 & n   &     &    \\ 
513 & \nodata & B2V        & 20 33 22.49 & +41 22 16.9 & 4.9 & 14.26 & -2.50    & 1.42 & -0.21 & 10.0 & 10.0 & 11.8 & n   &     &    \\ 
515 & B1V & B1V            & 20 33 23.24 & +41 13 41.9 & 6.8 & 14.66 & -3.20    & 2.03 & -0.23 & 13.8 & 13.8 & 11.0 &     &     &    \\
516 & O5.5V & O5.5V        & 20 33 23.46 & +41 09 12.9 & 7.5 & 11.84 & -5.07    & 2.20 & -0.30 & 34.2 & 34.2 &  9.5 &     &     &    \\
517 & \nodata & B1V        & 20 33 23.37 & +41 20 17.2 & 5.2 & 13.74 & -3.20    & 1.50 & -0.23 & 13.8 & 13.8 & 11.7 & n   &     &    \\ 
522 & \nodata & B2V(e?)    & 20 33 24.78 & +41 22 04.5 & 4.8 & 14.06 & -2.50    & 1.38 & -0.21 & 10.0 & 10.0 & 11.7 & n   & BCEP; $P=0.21210d?$    & 2   \\ 
531 & O8.5V & O8.5V        & 20 33 29.42 & +41 21 54.1 & 5.6 & 11.58 & -4.19    & 1.57 & -0.28 & 19.8 & 19.8 & 10.2 &     &     &    \\ 
534 & O7.5V & O8.5V        & 20 33 26.77 & +41 10 59.5 & 6.5 & 13.00 & -4.19    & 1.87 & -0.28 & 19.8 & 19.8 & 10.7 &     &     &    \\ 
539 & \nodata & B(5?)Ve    & 20 33 27.21 & +41 35 57.8 & 7.3 & 14.61 & -1.35    & 2.29 & -0.15 &  5.9 &  5.9 &  7.3 & n   &     &    \\ 
554 & \nodata & B4Ve       & 20 33 30.55 & +41 20 17.3 & 4.7 & 14.41 & -1.65    & 1.41 & -0.16 &  6.4 &  6.4 &  9.7 & n   & EA  & 2  \\ 
555 & O8V/ & \nodata       & 20 33 30.43 & +41 35 57.5 & 6.6 & 12.51 & -4.34    & 1.90 & -0.28 & 22.0 & 22.0 & 10.2 &     &     &    \\ 
556 & B1I & B1I            & 20 33 30.81 & +41 15 22.7 & 5.9 & 11.01 & -6.50    & 1.77 & -0.19 & 24.0 & 26.0 & 11.6 &     & Irr & 2   \\ 
561 & \nodata & B2V        & 20 33 31.68 & +41 21 46.1 & 4.5 & 13.73 & -2.50    & 1.30 & -0.21 & 10.0 & 10.0 & 11.7 & n   &     &    \\ 
568 & \nodata & B3V        & 20 33 33.38 & +41 08 36.3 & 7.0 & 14.76 & -1.90    & 2.16 & -0.18 &  7.6 &  7.6 &  9.6 & n   &     &    \\ 
573 & \nodata & B3I        & 20 33 33.97 & +41 19 38.4 & 5.3 & 13.87 & -6.30    & 1.62 & -0.13 & 22.0 & 23.5 & 14.9 & n   &     &    \\ 
575 & B2V & B2Ve           & 20 33 34.36 & +41 18 11.6 & 5.9 & 13.41 & -2.50    & 1.77 & -0.21 & 10.0 & 10.0 &  9.9 &     &     &    \\ 
576 & \nodata & B(7?)V     & 20 33 34.60 & +41 21 37.4 & 4.6 & 14.71 & -1.00    & 1.39 & -0.13 &  4.6 &  4.6 & 10.1 & n   &     &    \\ 
588 & B0V & B0V            & 20 33 37.02 & +41 16 11.4 & 5.8 & 12.40 & -3.80    & 1.66 & -0.26 & 17.5 & 17.5 & 10.4 &     & Cst & 2   \\ 
601 & O9.5III & B0Iab      & 20 33 39.14 & +41 19 26.1 & 5.1 & 11.07 & -6.50    & 1.47 & -0.22 & 25.0 & 27.0 & 12.5 &     & IS: & 3  \\ 
605 & B1V & B1V            & 20 33 39.84 & +41 22 52.4 & 4.3 & 11.78 & -3.20    & 1.19 & -0.23 & 13.8 & 13.8 & 10.7 &     &     &    \\ 
611 & O7V & O7V            & 20 33 40.88 & +41 30 18.5 & 5.5 & 12.77 & -4.63    & 1.55 & -0.29 & 26.5 & 26.5 & 11.8 &     &     &    \\ 
620 & \nodata & B0V        & 20 33 42.38 & +41 11 45.8 & 6.2 & 13.89 & -3.80    & 1.82 & -0.26 & 17.5 & 17.5 & 11.4 & n   &     &    \\ 
621 & \nodata & B1(V?)     & 20 33 42.57 & +41 14 56.9 & 6.3 & 14.93 & -3.20    & 1.91 & -0.18 & 13.8 & 13.8 & 11.8 & n   &     &    \\ 
632 & O9I & O9I            & 20 33 46.15 & +41 33 00.5 & 5.6 &  9.88 & -6.29    & 1.59 & -0.27 & 32.0 & 34.5 & 10.5 &     &     &    \\ 
635 & \nodata & B1III      & 20 33 46.85 & +41 08 01.9 & 5.8 & 13.81 & -4.30    & 1.72 & -0.21 & 17.5 & 18.3 & 12.3 & n   &     &    \\ 
639 & \nodata & B2V        & 20 33 47.63 & +41 09 06.5 & 6.0 & 14.37 & -2.50    & 1.77 & -0.21 & 10.0 & 10.0 & 10.9 & n   &     &    \\ 
641 & \nodata & B5(V?)     & 20 33 47.58 & +41 29 57.7 & 5.0 & 14.27 & -1.35    & 1.51 & -0.15 &  5.9 &  5.9 &  9.3 & n   &     &    \\ 
642 & B1III & B1III        & 20 33 47.88 & +41 20 41.7 & 5.3 & 11.78 & -4.30    & 1.55 & -0.21 & 17.5 & 18.3 & 10.8 &     &     &    \\ 
645 & \nodata & B2III      & 20 33 48.40 & +41 13 14.1 & 6.2 & 14.65 & -3.70    & 1.87 & -0.19 & 14.8 & 15.0 & 12.1 & n   &     &    \\ 
646 & B(1.5?)V & B1.5V     & 20 33 48.88 & +41 19 40.9 & 4.8 & 13.34 & -2.80    & 1.39 & -0.22 & 11.9 & 11.9 & 11.3 &     &     &    \\ 
650 & \nodata & B2V(e?)    & 20 33 48.83 & +41 37 39.7 & 5.2 & 14.94 & -2.50    & 1.53 & -0.21 & 10.0 & 10.0 & 12.2 & n   &     &    \\ 
692 & B0V & B0V            & 20 33 59.32 & +41 05 38.4 & 5.9 & 13.61 & -3.80    & 1.69 & -0.26 & 17.5 & 17.5 & 11.5 &     &     &    \\ 
696 & O9.5V & O9.5V        & 20 33 59.57 & +41 17 36.1 & 5.8 & 12.32 & -3.90    & 1.65 & -0.27 & 16.5 & 16.5 & 10.4 &     & EW/KE; $P=1.46d$ & 4   \\ 
712 & \nodata & B1V        & 20 34 04.43 & +41 08 08.4 & 6.1 & 13.66 & -3.20    & 1.80 & -0.23 & 13.8 & 13.8 & 10.7 & n   &     &    \\ 
716 & O9V & O9V            & 20 34 04.95 & +41 05 13.2 & 6.4 & 13.50 & -4.05    & 1.84 & -0.28 & 18.0 & 18.0 & 11.1 &     &     &    \\ 
720 & B? & O9.5V           & 20 34 06.10 & +41 08 09.6 & 7.0 & 13.59 & -3.90    & 2.05 & -0.27 & 16.5 & 16.5 & 10.5 &     &     &    \\ 
734 & O5I & O5I            & 20 34 08.54 & +41 36 59.3 & 5.4 & 10.03 & -6.33    & 1.49 & -0.32 & 50.9 & 53.0 & 10.9 &     & Susp & 1   \\ 
736 & O9V & O9V            & 20 34 09.52 & +41 34 13.4 & 5.2 & 12.79 & -4.05    & 1.46 & -0.28 & 18.0 & 18.0 & 11.6 &     &     &    \\ 
745 & O7V & O7V            & 20 34 13.50 & +41 35 02.6 & 5.4 & 11.91 & -4.63    & 1.50 & -0.29 & 26.5 & 26.5 & 11.1 &     &     &    \\ 
759 & \nodata & B1V        & 20 34 24.56 & +41 26 24.7 & 5.7 & 14.65 & -3.20    & 1.67 & -0.23 & 13.8 & 13.8 & 12.1 & n   &     &    \\ 
771 & O7V & O7V            & 20 34 29.52 & +41 31 45.5 & 7.0 & 12.06 & -4.63    & 2.05 & -0.29 & 26.5 & 26.5 &  9.6 &     &     &    \\ 
793 & B1.5III & B2IIIe     & 20 34 43.51 & +41 29 04.8 & 5.2 & 12.29 & -3.70    & 1.54 & -0.19 & 14.8 & 15.0 & 10.7 &     &     &    \\ 
\enddata 
\tablecomments{1. Star name in \mt\ nomenclature; 2. Spectral type obtained from the
literature; 3. Spectral type as determined by this survey; 4. Right Ascension,
(J2000); 5. Declination (J2000); 6. Calculated visual extinction in
magnitudes using \mt\ colors, \citet{Weg94} intrinsic colors, and
$R_V=$3.0; 7. Apparent magnitude as measured by \mt; 8.  Absolute
magnitudes from \citet{HM84} (B stars) and \citet{FM05} (O stars);
9. Measured colors from \mt; 10. Intrinsic colors from \citet{Weg94};
11. Initial masses interpolated from the stellar evolutionary models of
\citet{schaerer}; 12.  Current masses from \citet{FM05} and \citet{drilling};
 13. True distance moduli; 14. Notes: An 'n' denotes new
early type classification. A 'b' indicates a close visual double, 
unresolved in our data; 15. Photometric activity as found in the 
literature; 16. Literature reference for photometric activity.}

\tablerefs{(1) \citet{kukar81}; (2) \citet{PJ98}; 
 (3) \citet{Romano69}; (4) \citet{Rios2004}}
\end{deluxetable}

\clearpage
\pagestyle{plaintop}

\begin{deluxetable}{lrrrrrrrrr}
\tabletypesize{\tiny}
\tabletypesize{\scriptsize}
\tablewidth{35pc}
\tablecaption{Non OB Stars \label{nonOB.tab}}
\tablehead{
\colhead{Star} & 
\colhead{RA} & 
\colhead{Dec} & 
\colhead{V} & 
\colhead{S.C.}  \\ 
\colhead{(MT)} & 
\colhead{(J2000)} & 
\colhead{(J2000)} & 
\colhead{} & 
\colhead{} \\
\colhead{(1)} & 
\colhead{(2)} & 
\colhead{(3)} & 
\colhead{(4)} & 
\colhead{(5)}}
 \startdata
 34   & 20 30 54.41  & +41 32 49.3  & 15.44 & G?         \\
 35   & 20 30 54.94  & +41 35 45.6  & 15.63 & G-K        \\
 50   & 20 31 05.61  & +41 34 30.7  & 14.33 & G          \\
 52   & 20 31 08.25  & +41 35 32.3  & 10.98 & F          \\
 54   & 20 31 08.33  & +41 35 37.1  & 11.26 & G          \\
 62   & 20 31 11.79  & +41 34 24.5  & 14.76 & G          \\
 100  & 20 31 32.44  & +41 37 41.3  & 15.60 & G          \\
 107  & 20 31 34.05  & +41 31 02.6  & 14.80 & G?         \\
 118  & 20 31 38.59  & +41 19 53.8  & 14.29 & F?         \\
 126  & 20 31 41.49  & +41 23 03.7  & 15.02 & G          \\
 133  & 20 31 42.68  & +41 29 54.5  & 15.04 & G?         \\
 137  & 20 31 44.89  & +41 21 38.1  & 14.47 & G          \\
 141  & 20 31 45.91  & +41 34 49.2  & 15.36 & G          \\
 156  & 20 31 52.64  & +41 25 35.8  & 15.37 & G          \\
 235  & 20 32 19.81  & +41 23 51.5  & 13.99 & G          \\
 242  & 20 32 23.57  & +41 19 24.3  & 15.36 & G          \\
 244  & 20 32 23.83  & +41 19 36.2  & 15.08 & A          \\
 267  & 20 32 31.45  & +41 14 08.8  & 12.87 & G          \\
 272  & 20 32 32.39  & +41 32 37.3  & 15.01 & G          \\
 279  & 20 32 34.33  & +41 04 26.0  & 13.85 & A          \\
 289  & 20 32 35.79  & +41 35 37.8  & 14.71 & A          \\
 308  & 20 32 41.37  & +41 30 26.7  & 15.12 & G          \\
 398  & 20 33 04.40  & +41 32 55.8  & 14.70 & G          \\
 430  & 20 33 10.76  & +41 07 20.5  & 14.41 & G          \\
 436  & 20 33 11.07  & +41 14 45.6  & 16.08 & K          \\
 438  & 20 33 10.69  & +41 39 52.9  & 14.75 & K          \\
 463  & 20 33 14.86  & +41 19 34.7  & 15.01 & F          \\
 519  & 20 33 23.85  & +41 30 37.8  & 15.06 & G          \\
 524  & 20 33 24.70  & +41 40 59.3  & 13.06 & A          \\
 526  & 20 33 25.37  & +41 30 28.1  & 14.59 & M          \\
 617  & 20 33 42.15  & +41 22 22.8  & 14.93 & late?      \\
 637  & 20 33 46.58  & +41 38 34.3  & 15.00 & G          \\
 665  & 20 33 53.65  & +41 25 32.3  & 14.76 & K          \\
 728  & 20 34 06.64  & +41 43 13.3  & 15.15 & K          \\
 735  & 20 34 09.47  & +41 29 43.9  & 15.38 & F          \\
 737  & 20 34 09.85  & +41 25 57.5  & 14.74 & G          \\
 789  & 20 34 39.49  & +41 37 46.2  & 11.63 & late       \\
\enddata
\tablecomments{1.Star name in \mt\
nomenclature; 2. Right Ascension (J2000); 3. Declination (J2000); 
4. Apparent magnitude as measured by \mt; 5.
Spectral type.}
\end{deluxetable}

\clearpage

\begin{deluxetable}{llll}
\tabletypesize{\scriptsize}
\tablecaption{Complete Table of Observations and Radial Velocities\label{electr.tab}}
\tablewidth{0pt}
\tablehead{
\colhead{Star} &
\colhead{HJD} &
\colhead{Rad. Vel.} &
\colhead{$\sigma_v$} \\
\colhead{(MT)} &
\colhead{} &
\colhead{(\kms)} &
\colhead{(\kms)} \\
\colhead{(1)} &
\colhead{(2)} &
\colhead{(3)} &
\colhead{(4)}}
\startdata
020 &	2451365.033 & -10.9 &  4.5 \\
020 &	2451467.752 & -11.7 &  3.9 \\
020 &	2451805.998 & -18.5 &  4.5 \\
\nodata & \nodata & \nodata & \nodata
\enddata
\tablecomments{1. Star name in \mt\ nomenclature; 2. Heliocentric Julian date;
3. Heliocentric radial velocity; 4. 1$\sigma$ uncertainty.   The full contents of 
Table~\ref{electr.tab} are available in machine-readable form in the 
electronic edition of the {\it Astrophysical Journal}. }
\end{deluxetable}

\clearpage

\begin{deluxetable}{lrrrrrrrrc}
\tabletypesize{\scriptsize}
\tablecaption{Derived Radial Velocity Parameters \label{veltable.tab}}
\tablewidth{0pt}
\tablehead{
\colhead{Star} &
\colhead{S.C.} &
\colhead{$\mathrm{N}_{\mathrm{obs}}$} &
\colhead{$\mathrm{V}_{\mathrm{avg}}$} &
\colhead{$\mathrm{V}_{\mathrm{mid}}$} &
\colhead{$\mathrm{V}_{\mathrm{h}}$} &
\colhead{$\mathrm{V}_{\mathrm{rms}}$} &
\colhead{$\overline{\sigma}_{v}$} &
\colhead{P$(\chi^2,\nu)$} &
\colhead{Sp. Activ.} \\
\colhead{(MT)} &
\colhead{} &
\colhead{} &
\colhead{(\kms)} &
\colhead{(\kms)} &
\colhead{(\kms)} &
\colhead{(\kms)} &
\colhead{(\kms)} &
\colhead{} &
\colhead{}  \\
\colhead{(1)} &
\colhead{(2)} &
\colhead{(3)} &
\colhead{(4)} &
\colhead{(5)} &
\colhead{(6)} &
\colhead{(7)} &
\colhead{(8)} &
\colhead{(9)} &
\colhead{(10)}}
\startdata
20 & B0V &          8 & -13.1  & -22.2  &  12.2  &  9.1   &  6.9   &  0.607    &      \\
21 & B2II &         4 & -40.8  & -38.4  &  17.6  &  16.0  &  7.1   &  0.000    & SB1  \\
59 & O8V &         17 & -10.1  & -29.1  &  77.3  &  54.1  &  11.0  &  0.000    & SB1  \\
70 & O9V &          7 & -6.4   & -15.1  &  12.6  &  8.8   &  4.9   &  0.102    &      \\
83 & B1I &         12 & -4.7   & -4.8   &  3.7   &  2.3   &  1.8   &  0.000    & SB1  \\
97 & B2V &          7 & -4.4   & -8.9   &  16.8  &  13.3  &  11.0  &  0.306    &      \\
103 & B1V &         5 & -26.4  & -15.9  &  18.3  &  14.6  &  9.8   &  0.382    &      \\
106 & B3V &         6 & -6.2   & -9.4   &  13.7  &  8.9   &  7.9   &  0.111    &      \\
129 & B3V &         6 & -12.3  & -15.1  &  13.6  &  9.7   &  12.3  &  0.742    &      \\
138 & O8I &        15 & -7.7   &  19.7  &  58.2  &  31.2  &  8.7   &  0.000    & SB1  \\
145 & O9III &      18 & -12.4  & -22.7  &  41.1  &  28.4  &  3.5   &  0.000    & SB1  \\
164 & B3V &         3 & -8.6   & -13.7  &  11.4  &  11.6  &  14.0  &  0.695    &      \\
169 & B2V &         3 & -24.5  & -27.6  &  10.5  &  11.7  &  7.8   &  0.179    &      \\
174 & B2IV &        6 & -17.5  & -16.5  &  6.6   &  5.8   &  3.1   &  0.001    & SB1    \\
187 & B1V &         5 & -5.1   & -3.8   &  3.6   &  2.8   &  5.3   &  0.971    &      \\
191 & B3IV &        7 & -10.4  & -9.5   &  5.5   &  3.7   &  5.2   &  0.942    &      \\
196 & B6V &         3 & -34.2  & -34.5  &  37.8  &  41.9  &  16.2  &  0.001    & SB1     \\
200 & B3V &         6 & -6.1   &  4.8   &  15.9  &  11.6  &  12.1  &  0.915    &      \\
202 & B2V &         5 & -2.0   &  15.9  &  34.4  &  27.0  &  13.1  &  0.000    & SB1     \\
213 & B0V &        10 &  2.8   &  11.2  &  27.8  &  19.7  &  13.3  &  0.215    &      \\
215 & B2V &         3 & -2.8   & -1.4   &  3.1   &  3.3   &  14.0  &  0.988    &      \\
216 & B1.5V &       4 & -14.7  & -13.5  &  5.4   &  5.7   &  7.4   &  0.785    &      \\
217 & O7IIIf &     12 & -6.5   & -5.9   &  4.4   &  2.6   &  5.0   &  0.884    &      \\
220 & B1V &         6 & -11.2  & -6.9   &  13.2  &  10.6  &  7.8   &  0.073    &      \\
227 & O9V &        14 & -8.4   & -19.1  &  30.5  &  14.9  &  10.0  &  0.416    &      \\
234 & B2V &         7 & -16.2  & -11.8  &  17.8  &  12.1  &  5.2   &  0.000    & SB1     \\
238 & B1V &         4 & -7.5   & -4.4   &  30.6  &  25.3  &  14.7  &  0.029    & SB1?     \\
239 & B4V &         7 &  2.5   & -1.1   &  13.7  &  10.1  &  13.3  &  0.851    &      \\
241 & B2V &         5 & -6.3   & -4.7   &  13.6  &  10.3  &  4.8   &  0.001    & SB1     \\
248 & B2V &         5 & -12.7  & -22.7  &  12.8  &  10.5  &  9.0   &  0.456    &       \\
250 & B2III &       5 & -8.8   & -9.9   &  3.8   &  3.0   &  3.3   &  0.598    &      \\
252 & B1.5II &      8 & -3.7   &  2.7   &  32.7  &  19.5  &  7.6   &  0.000    & SB1/SB2     \\
255 & B2III &       7 & -12.9  & -13.2  &  10.1  &  8.2   &  7.6   &  0.412    &      \\
258 & O8V &        17 & -20.1  &  2.6   &  64.0  &  34.6  &  10.2  &  0.000    & SB1     \\
259 & B0Ib &       11 & -14.2  & -19.0  &  10.4  &  5.6   &  3.5   &  0.014    & SB1?     \\
264 & B2III &       7 & -1.7   & -1.7   &  5.2   &  3.7   &  6.6   &  0.831    &      \\
268 & B2.5V &       7 & -26.3  & -22.5  &  28.6  &  23.6  &  8.8   &  0.000    & SB1     \\
271 & B4V &         6 & -31.1  & -30.3  &  22.2  &  16.5  &  14.9  &  0.268    &      \\
275 & B2V &         4 & -18.6  & -19.9  &  7.5   &  6.3   &  8.3   &  0.879    &      \\
292 & B2V &         7 & -12.9  & -4.0   &  22.4  &  15.8  &  5.6   &  0.000    & SB1/SB2?     \\
295 & B2V &         4 & -5.2   & -5.9   &  2.4   &  2.0   &  6.6   &  0.992    &      \\
298 & B3V &         7 & -23.8  &  3.0   &  40.5  &  31.3  &  17.8  &  0.029    & SB1?     \\
299 & O7V &        16 & -15.7  & -14.2  &  24.1  &  12.5  &  9.9   &  0.214    &      \\
300 & B1V &         6 & -12.7  & -11.9  &  5.7   &  5.2   &  4.4   &  0.207    &      \\
311 & B2V &         6 &  12.2  &  5.9   &  28.8  &  22.7  &  9.0   &  0.000    & SB1     \\
317 & O8V &        13 & -11.64 &  3.6   &  19.7  &  10.9  &  10.1  &  0.834    &      \\
322 & B2.5V &       6 & -27.3  & -21.7  &  17.1  &  12.7  &  11.0  &  0.512    &      \\
325 & B1.5II &      6 & -8.2   & -9.0   &  4.3   &  3.1   &  4.4   &  0.843    &      \\
336 & B3III &       7 & -12.8  & -12.4  &  14.4  &  10.2  &  5.5   &  0.001    & SB1     \\
339 & O8V &        13 & -15.1  & -15.3  &  24.4  &  11.8  &  6.9   &  0.299    &      \\
343 & B1V &         7 & -5.2   & -6.5   &  12.3  &  10.1  &  9.6   &  0.571    &      \\
365 & B1V &         6 & -12.1  & -12.9  &  11.5  &  7.6   &  7.5   &  0.759    & SB2?     \\
372 & B0V &         3 & -5.7   & -3.8   &  63.2  &  63.7  &  11.5  &  0.000    & SB1     \\
376 & O8V &        11 & -14.0  & -26.8  &  19.0  &  11.2  &  9.3   &  0.716    &        \\
378 & B0V &         7 & -21.9  & -4.9   &  38.6  &  25.9  &  11.1  &  0.001    & SB1/SB2?     \\
390 & O8V &         6 & -16.7  & -20.0  &  11.6  &  8.3   &  9.1   &  0.567    &      \\
395 & B1V &         8 & -14.6  & -17.4  &  8.3   &  5.1   &  2.8   &  0.047    &      \\
400 & B1V &         6 & -9.8   & -9.1   &  5.3   &  4.0   &  5.7   &  0.775    &      \\
403 & B1V &         8 & -7.0   & -18.1  &  41.2  &  33.9  &  8.2   &  0.000    & SB1     \\
409 & B0.5V &       7 & -15.5  & -11.3  &  13.9  &  11.1  &  10.6  &  0.428    &      \\
428 & B1V &         7 & -2.4   & -5.1   &  12.4  &  8.5   &  4.7   &  0.000    & SB1     \\
429 & B0V &         8 & -15.9  &  8.6   &  34.7  &  23.2  &  5.3   &  0.000    & SB1  \\
431 & O5If &       13 & -16.1  & -14.7  &  48.0  &  23.4  &  13.0  &  0.083    &      \\
435 & B0V &         5 & -10.3  & -9.9   &  11.0  &  8.8   &  11.2  &  0.633    &      \\
441 & B2III? &      5 & -9.3   & -8.1   &  4.4   &  3.8   &  6.1   &  0.908    &      \\
448 & O6V &         4 & -33.0  & -5.0   &  39.3  &  37.0  &  13.8  &  0.003    & SB1 \\
453 & B5?V &        4 & 0.1    & -10.8  &  15.7  &  14.2  &  9.3   &  0.299    &      \\
457 & O3If &       11 & -27.6  & -28.1  &  14.1  &  8.4   &  8.2   &  0.570    &      \\
462 & O7III &      13 & -11.3  & -2.3   &  19.4  &  9.6   &  7.6   &  0.452    &      \\
465 & O5.5I &      17 & -10.6  & -11.5  &  16.5  &  9.6   &  6.5   &  0.000    & SB1/SB2\tablenotemark{a}  \\
467 & B1V &         5 & -10.2  & -6.8   &  13.3  &  11.5  &  10.7  &  0.551    &      \\
469 & B1III &       5 & -10.6  & -10.5  &  3.1   &  2.5   &  3.3   &  0.851    &      \\
470 & O9V &         6 & -19.8  & -30.2  &  18.1  &  13.4  &  7.9   &  0.226    &      \\
473 & O8.5V &      10 & -5.5   & -9.4   &  35.5  &  22.5  &  13.3  &  0.208    &      \\
477 & B0V &         6 & -10.2  & -13.4  &  17.1  &  12.9  &  17.7  &  0.753    &      \\
480 & O7V &        11 & -12.6  & -13.5  &  21.3  &  12.0  &  13.5  &  0.595    &      \\
483 & O5III &      12 & -15.5  & -12.6  &  22.7  &  11.5  &  8.2   &  0.025    & SB1?     \\
485 & O8V &         8 & -11.9  & -34.7  &  29.3  &  18.7  &  7.8   &  0.020    & SB1?     \\
490 & B0? &         5 & -10.9  &  5.4   &  39.7  &  29.2  &  19.6  &  0.112    &      \\
492 & B1V &         4 &  14.6  &  23.5  &  40.8  &  36.0  &  14.8  &  0.000    & SB1     \\
493 & B5IV &        4 & -5.1   &  3.1   &  57.2  &  48.7  &  13.7  &  0.000    & SB1     \\
507 & O9V &         5 & -16.6  & -19.5  &  6.3   &  5.1   &  7.1   &  0.679    &      \\
509 & B0III &       4 & -13.4  & -11.6  &  3.3   &  2.9   &  4.7   &  0.964    &      \\
513 & B2V &         5 & -11.6  & -35.0  &  36.7  &  31.3  &  7.3   &  0.000    & SB1    \\
515 & B1V &         7 & -7.3   & -9.6   &  12.4  &  7.5   &  6.5   &  0.559    &      \\
516 & O5.5V &      11 & -8.3   &  4.7   &  56.6  &  26.6  &  14.1  &  0.115    &      \\
517 & B1V &         6 & -3.4   & -5.1   &  12.0  &  10.1  &  5.9   &  0.014    & SB1?     \\
522 & B2Ve? &       7 & 0.6    &  25.1  &  60.9  &  41.3  &  13.5  &  0.000    & SB1  \\
531 & O8.5V &       6 & -14.8  & -17.0  &  7.9   &  5.4   &  4.0   &  0.171    &      \\ 
534 & O8.5V &       6 & -13.8  & -14.6  &  6.6   &  4.6   &  6.4   &  0.632    &      \\
555 & O8V &         7 & -6.5   & -9.6   &  19.4  &  14.6  &  7.6   &  0.002    & SB1     \\
556 & B1I &        14 & -6.3   & 0.2    &  14.0  &  7.1   &  3.7   &  0.000    & SB1     \\
561 & B2V &         6 & -15.4  & -19.0  &  23.8  &  19.8  &  6.8   &  0.000    & SB1/SB2?     \\
568 & B3V &         7 & -14.1  & -11.2  &  26.2  &  19.6  &  13.2  &  0.042    &      \\
573 & B3I &         7 & -10.7  & -9.7   &  7.6   &  6.7   &  5.1   &  0.010    & SB1?     \\
576 & B7?V &        5 & -1.4   & -4.5   &  5.5   &  4.5   &  10.3  &  0.979    &      \\
588 & B0V &         5 & -4.9   & -16.5  &  22.1  &  17.6  &  9.6   &  0.062    &      \\
601 & B0Iab &       8 &  3.0   & -2.6   &  15.6  &  9.5   &  5.1   &  0.000    & SB1     \\
605 & B1V &        11 & -11.0  & -20.2  &  26.2  &  13.8  &  9.0   &  0.470    & SB2?     \\
611 & O7V &         5 & -22.7  & -23.1  &  1.4   &  1.2   &  4.5   &  0.998    &      \\
620 & B0V &         6 & -11.2  & -8.6   &  14.9  &  10.1  &  5.7   &  0.031    & SB1?     \\
621 & B1V? &        5 & -4.9   & -10.0  &  16.3  &  12.1  &  10.8  &  0.695    &      \\
632 & O9I &        14 & -2.3   & -9.8   &  13.6  &  6.7   &  5.6   &  0.253    &      \\
635 & B1III &       8 & -8.2   & -14.0  &  8.5   &  5.4   &  2.3   &  0.111    &      \\
639 & B2V &         4 & -3.1   &  25.2  &  44.2  &  40.5  &  17.8  &  0.104    & SB2?     \\
641 & B5V? &        7 & -13.8  & -12.0  &  12.0  &  7.1   &  14.7  &  0.995    &      \\
642 & B1III &      15 & -8.7   & -26.1  &  39.2  &  20.1  &  7.7   &  0.000    & SB1     \\
645 & B2III &       6 & -10.1  & -7.3   &  5.4   &  4.2   &  4.2   &  0.233    &      \\
646 & B1.5V &       5 & -7.9   & -11.3  &  9.7   &  7.6   &  6.1   &  0.300    &      \\
650 & B2Ve? &       5 & -8.6   & -16.3  &  14.6  &  11.3  &  12.9  &  0.792    &      \\
692 & B0V &         6 & -4.2   & -6.3   &  10.4  &  8.3   &  9.6   &  0.778    & SB2?     \\
696 & O9.5V &       9 & 0.4    & -6.9   &  49.7  &  33.3  &  16.2  &  0.000    & SB1/SB2\tablenotemark{b} \\
712 & B1V &         5 & -13.8  & -1.7   &  20.3  &  16.4  &  13.6  &  0.313    &      \\
716 & O9V &         5 & -11.5  & -14.8  &  7.9   &  6.2   &  4.5   &  0.351    &      \\
720 & O9.5V &       5 & -6.9   & -17.2  &  90.6  &  67.7  &  20.3  &  0.000    & SB1/SB2    \\
734 & O5I &        15 & -27.5  & -11.8  &  39.2  &  23.5  &  6.7   &  0.000    & SB1     \\
736 & O9V &         5 & -12.3  & -13.0  &  4.0   &  3.2   &  6.0   &  0.909    &      \\
745 & O7V &        12 & -15.8  & -39.5  &  50.4  &  24.7  &  11.6  &  0.035    & SB1?     \\
759 & B1V &         6 & -14.5  & -15.6  &  6.4   &  5.3   &  6.4   &  0.794    & SB2?     \\
771 & O7V &        10 & -8.1   & -14.6  &  69.1  &  35.0  &  13.4  &  0.000    & SB1/SB2     
\enddata
\tablecomments{1. Star name in \mt\ nomenclature; 2. Spectral
types as determined/accepted by this survey; 3. Number of
usable observations; 4. The weighted average heliocentric velocity;
5. $V_{mid}\equiv 0.5 (V_{max} + V_{min})$, the simple average of the
largest and smallest observed heliocentric velocity; 6. $V_h=0.5
(V_{max} - V_{min})$, a measure of the velocity semi-amplitude; 7. The
RMS heliocentric velocity dispersion of all observations; 8. The mean 
velocity uncertainty, averaged over all observations; 9. The 
probability (P) that $\chi^2$ would be exceeded by 
chance, given $\nu=N_{obs}-1$ degrees of freedom; 10. Spectroscopic 
activity where ``SB1'' represents probable single-lined 
variability ($P(\chi^2,\nu)\leq0.01$), ``SB1'' represents possible 
single-lined variability ($0.01<P(\chi^2,\nu)\leq0.04$), ``SB2?'' 
represents a possible double-lined binary signature, and ``SB2'' is a definite
double-lined signature.}

\tablenotetext{a}{Spectroscopic period of $P=21.908d$ \citep{Debeck2004}.}
\tablenotetext{b}{Photometric period of $P=1.46d$ \citep{Rios2004}.}

\end{deluxetable}

\clearpage

\clearpage

\clearpage

\clearpage

\clearpage

\clearpage

\clearpage

\clearpage

\clearpage

\clearpage

\clearpage

\clearpage

\clearpage

\clearpage

\clearpage

\clearpage

\clearpage

\clearpage

\clearpage

\clearpage

\clearpage

\clearpage

\clearpage

\end{document}